**Multimodal Fusion and Interpretability in Human Activity Recognition:**
**A Reproducible Framework for Sensor-Based Modeling**

Authors (ORCID): Yiyao Yang[1] (0009-0001-8693-4888), Yasemin Gulbahar[2] (0000-0002-1726-3224)

Emails: {yy3555[1], yg2918[2]}@tc.columbia.edu

Affiliation: Columbia University, New York, NY, United States

Corresponding Author: Yiyao Yang (yy3555@tc.columbia.edu)

Author Contributions:
Conceptualization: Yasemin Gulbahar, Yiyao Yang
Data curation: Yasemin Gulbahar
Formal analysis: Yiyao Yang
Funding acquisition: Not applicable
Investigation: Yiyao Yang
Methodology: Yiyao Yang
Formal analysis and investigation: Yiyao Yang
Project administration: Yiyao Yang, Yasemin Gulbahar
Resources: Yiyao Yang
Software: Yiyao Yang
Supervision: Yasemin Gulbahar
Validation: Yiyao Yang, Yasemin Gulbahar
Visualization: Yiyao Yang
Writing - original draft: Yiyao Yang
Writing - review and editing: Yiyao Yang

Acknowledgements: The authors have no acknowledgements to declare.

Competing Interests: The authors declare that they have no competing interests.

## I. Abstract

The research introduces a reproducible framework for transforming raw, heterogeneous sensor data streams into aligned, semantically meaningful representations for multimodal human activity recognition. Grounded in the Carnegie Mellon University Multi-Modal Activity Database (CMU-MMAC) database and focused on the naturalistic Subject 07–Brownie session, the study traces the full pipeline from data ingestion to modeling and interpretation. Unlike a black box preprocessing, a unified preprocessing workflow is proposed that temporally aligns video, audio, and Radio Frequency Identification (RFID) through resampling, grayscale conversion, sliding-window segmentation, and modality-specific normalization, producing standardized fused tensors suitable for downstream learning. Building on this foundation, the work systematically compares early, late, and hybrid fusion strategies using Long Short-Term Memory (LSTM)-based models implemented with PyTorch and TensorFlow, showing that late fusion consistently achieves the highest validation accuracy, with hybrid fusion outperforming early fusion. To evaluate interpretability and modality contribution, Principal Component Analysis (PCA) and t-distributed Stochastic Neighbor Embedding (t-SNE) visualizations reveal coherent temporal structure and confirm that the video carries stronger discriminative power than audio, while their combination yields substantial performance gains. Incorporating sparse, asynchronous RFID signals further improves accuracy by over 50% and boosts macro-averaged Receiver Operating Characteristic – Area Under the Curve (ROC-AUC), demonstrating the added value of object interaction cues. Overall, the framework contributes a modular, empirically validated approach to multimodal fusion that links preprocessing design, fusion architecture, and interpretability, offering a transferable template for intelligent systems operating in complex, real-world activity settings.

## III. Introduction

Multimodal data increasingly serve as valuable resources in advancing machine learning, human-computer interaction, and human activity recognition (HAR) research. In real-world settings where we observe behaviors and interactions, regular activities are subjected to complicated sensory dynamics, visual, auditory, and corporeal interactions that happen on diverse modalities and timescales. It does become imperative not only to exploit richly annotated sensor data but also powerful tools for transforming and integrating heterogeneous inputs into coherent, temporally aligned presentations so that we can model these situations correctly. Thus, through images, audio, and video transcriptions, it is possible to dive into the complex structure of human behavior.

However, multimodal integration entails a challenging methodology. Sensor streams typically are heterogeneous in terms of sampling rate, data format, signal structural nature, and semantic representation. Video arrives at a consistent 30 frames per second, audio is captured at kilohertz-scale resolution, and object interaction modes like RFID provide sparse, asynchronous events. Without careful preprocessing, these disparities lead to misalignment, modality imbalance, or information loss, compromising downstream model performance.

To address these challenges, this research presents a comprehensive, modular pipeline for the CMU Multi-Modal Activity Database (CMU-MMAC), a benchmark dataset for naturalistic human activity recognition. Unlike prior works, which narrow the focus to modeling or classifier performance, our work emphasizes the complete multimodal data life cycle from acquisition and conversion, alignment, through fusion, to interpretation. The research not only applies the framework to a prototypic sequence of cooking (Subject 07 – Brownie), but also explores a variety of fusion methods and RFID as a new, hitherto barely explored, but promising modality for introducing semantic richness.

By providing a thorough documentation of the various phases of the data preparation and modeling process of the multimodal data fusion, the present study offers a replicable methodology that links the theoretical underpinnings of the work to the practical implementation of the same within a naturalistic case setting. The study is presented as a demonstrative framework that may be extended to future work within various human activity recognition settings. In this manner, the study contributes to the transparency and replicability of the human activity recognition pipeline.

## IV. Multimodal Data for Exploring Human Activity

The multimodal human activity recognition (HAR) field has experienced great strides through fusing several sensor modalities, which have prompted rising lines of work directed towards critical challenges such as data fusion, data transformation, data frequency alignment, data format normalization, and data coding. Foundational work in the field has produced critical taxonomies and methodological orientations. For example, Aguileta et al. (2019) provided a definitive categorization of approaches towards sensor fusions, which exhibited sensor technique diversity and integration consistencies. Subsequent work, Vrigkas et al. (2015), concerned exploiting low-level features and high-level semantic knowledge towards improving recognition accuracy in various, realistic worlds. Accordingly, Aggarwal and Ryoo (2011) listed recognition systems based on space-time volume and hierarchical models, which emphasized sensor temporal structures and their influence towards capturing activities per segment.

Wearable sensor technology advances added further layers of complexity to HAR system design. Lara and Labrador (2013) tackled such complexity by advancing a two-level taxonomy for wearable HAR systems, which provides benchmarking approaches across diverse system architectures. As research shifted away from unimodal towards multimodal interaction paradigms, the contribution of fusion became more and more dominant. Jaimes and Sebe (2007), in their multimodal human-computer interaction survey, pinpointed data fusion as a perennial issue, notably in gesture, gaze, and affective state interpretation application domains. Though early, late, and hybrid paradigms of fusion are well known, Aguileta et al. (2019) noted significant variability in the implementation of those approaches. In a broader context, Baltrusaitis, Ahuja, and Morency (2019) provided one of the most comprehensive surveys and taxonomies of multimodal machine learning to date, identifying key challenges in representation, translation, alignment, fusion, and co-learning. The taxonomy served as a valuable reference point for the current study's focus on modular, reproducible fusion pipelines.

Variation in method across studies highlights a necessity for generalizable, reproducible, and integrative pipelines, particularly in sensor-rich, context-aware settings such as smart classrooms and domestic kitchens. The need for semantic richness in activity labels within such environments had also been addressed by Yordanova, Krüger, and Kirste (2018), who introduced a semantic annotation framework for the CMU-MMAC dataset. The method, based on plan operators, extended symbolic labeling by embedding contextual meaning, facilitating supervised learning and performance evaluation in naturalistic cooking environments.

Several studies had specifically leveraged the CMU-MMAC database to explore modality contributions and classification strategies. For instance, Fisher and Reddy (n.d.) utilized first-person video and IMU data to perform frame- and sequence-level classification for cooking tasks, employing an ensemble of classifiers including Support Vector Machines (SVMs), K-Nearest Neighbors (K-NNs), and Hidden Markov Models (HMMs). The work demonstrated the feasibility of combining egocentric and motion data for robust action recognition, aligning with broader trends in multimodal representation learning. Additionally, Macua (2012) offered a comprehensive technical overview of the CMU-MMAC database's structure, sensor setup, and organization, paving the way for more systematic data utilization in HAR research.

Efforts to address integration inconsistencies have led to the emergence of more advanced fusion architectures and temporal modeling strategies. Cai et al. (2023) investigated multi-sensor SLAM to support real-time path planning, demonstrating that precise spatial alignment enhances sensor coordination. Xie et al. (2025) extended this concept to HAR by developing a self-attention-based framework that combines skeleton and RGB inputs to enable more accurate temporal modeling. Further contributions include the work of Koutrintzes et al. (2022), who proposed a system for unifying video and skeletal data into 2D feature maps, streamlining feature learning across modalities. Wang et al. (2018) focused on temporal segmentation, introducing the Segment-Tube approach to localize actions within untrimmed video sequences using per-frame segmentation masks. Enhancing this trajectory of research, Feichtenhofer et al. (2019) introduced SlowFast networks, which operate on dual temporal streams, one slow for spatial semantics and one fast for high-frequency motion cues. The dual-pathway design proves especially effective in HAR contexts where data streams, such as video and motion capture, differ in sampling rate and resolution.

Simultaneously with architectural advances, deep learning has transformed approaches to cross-modal representation and alignment. Shaikh et al. (2024) and Guo et al. (2019) reviewed the transition of convolutional networks to Transformer-based models, noting attention mechanisms and encoder-decoder structures to be key to encoding modality-specific and shared features. As a complement to attention-centered fusion, Lindenbaum et al. (2020) and Katz et al. (2019) extended diffusion-based methods that operate to reduce cross-modal noise. Katz et al. (2019) specifically introduced alternating diffusion maps to retain shared latent structure and discard modality-specific artifacts. Subsequent work in graph-based models has yielded new lines of approaches to multimodal structure representation. Zhang et al. (2022) conducted a systematic review of deep learning approaches to graph-structured data, including graph convolutional networks and graph autoencoders. Such methods have strong prospects for encoding interaction between sensors or human joints in spatial-temporal HAR settings, and are a structured alternative to flat feature concatenation.

Expanding on these innovations, Zhang, Huang, Liu, and Sato (2024) proposed a masked autoencoder approach that jointly learns from egocentric video and body-worn IMU data using self-supervised pretraining. Their graph-based modeling of inter-IMU dynamics proves particularly robust in scenarios with missing devices or corrupted input. Deployment-focused work has also influenced research orientation, more so towards scalability and generalization to practical scenarios. Qi et al. (2018) also identified sensor-based HAR systems' necessity for reactivity and interoperability when they are deployed in Internet of Things (IoT) environments. Benos et al. (2024) also demonstrated that sensor placement and fusion strategy have a strong influence on recognition accuracy, at least in wearables. Ugonna et al. (2024) contributed further insight by also utilizing reinforcement learning and large-scale language models to create adaptive, intelligent fusion pipes. To improve generalizability across subjects and scenarios, Cao et al. (2025) also introduced the CLEAR model, which employs contrastive learning, data augmentation, and cross-modal representation learning. Similarly, Zhang et al. (2022) also introduced a CNN-GRU architecture that can scale to multi-scale fusion, balancing efficiency in computation with flexibility in modeling. In parallel, Cheng et al. (2022) addressed the need for temporally and spatially synchronized datasets by developing the HFUT-MMD, a large-scale multimodal dataset with precise sensor calibration. Their work demonstrates how low-precision modalities can be optimized to emulate high-quality motion capture, bolstering alignment fidelity across streams. Soran, Farhadi, and Shapiro (2015) also contributed to this line of inquiry by proposing a model that integrates egocentric and static camera views, dynamically weighting each input stream to maximize action recognition performance on the CMU-MMAC database.

Across the existing body of work, researchers increasingly agree upon principled methods of fusion and structured preprocessing pipelines. Nonetheless, there remain multiple unresolved challenges, chief among them being a dearth of standardized, general-purpose merging and preprocessing frameworks. The current study responds to this gap by introducing a reproducible, modular method for merging and harmonizing multimodal sensor data under naturalistic conditions, drawing empirical grounding from the CMU-MMAC database and extending prior work on both methodological reproducibility and modality integration. Hence, this research aims to develop and evaluate a novel multimodal data integration framework for identifying and classifying naturalistic human activities.

## V. Methodology

Using the open-source dataset of CMU-MMAC, this research employs a suite of quantitative methods to explore integration strategies for multimodal data. A combination of data mining techniques and statistical learning methods, including dimensionality reduction (PCA, t-SNE), classification models (LSTM, logistic regression, random forest), and evaluation metrics (accuracy, F1-score, ROC-AUC), is applied to assess the effectiveness, interpretability, and robustness of different fusion strategies. All preprocessing, modeling, and visualization tasks are conducted using Python and its scientific computing ecosystem, including libraries such as NumPy, pandas, scikit-learn, PyTorch, and matplotlib.

### (V.1) Data Source

De la Torre, F., Hodgins, J., Bargteil, A., Martin, X., Macey, J., Collado, A., & Vu, T. (2009). *CMU Multimodal Activity (Kitchen) Dataset* [Data set]. Carnegie Mellon University. https://kitchen.cs.cmu.edu/

### (V.2) Data Description

The Carnegie Mellon University Multi-Modal Activity Database (CMU-MMAC) database captures a rich multimodal dataset for measuring human activity in the multimodal environment of food preparation and cooking. These multimodal data are recorded at Carnegie Mellon's Motion Capture Lab with twenty-five users preparing a total of five dishes: brownies, pizza, sandwiches, salads, and scrambled eggs. The kitchen environment is carefully planned in a naturalistic but experimentally controllable environment, so that a complete multimodal observation becomes possible.

The data collection approach involves the use of varied sensor modalities. Video data are captured through six cameras: three high-definition color video cameras (1024 × 768) at 30 Hz, two low-definition video cameras (640 × 480) at 60 Hz, and a wearable high-definition video camera (800 × 600 or 1024 × 768) worn on the participant, operating at 30 Hz. Audio data are captured through five balanced condenser microphones placed at strategic locations around the kitchen for the capture of ambient noise, speech, and manipulations of utensils. Motion capture is captured through a Vicon setup of 12 MX-40 infrared cameras, operating at 120 Hz, each at 4-megapixel resolution to enable precise skeletal capture through reflective markers.

In addition to visual and acoustic modalities, inertial measurement units (IMUs) are used to track movement dynamics. These include both wired IMUs (3DMGX) and Bluetooth-enabled wireless IMUs (6DOF), worn on various body locations. Wearable devices such as the BodyMedia armband (measuring skin temperature, galvanic skin response, and energy expenditure) and the eWatch (recording ambient light, temperature, and motion) are also used to capture physiological and ambient context. Together, these sensors provide a comprehensive, temporally aligned, and high-dimensional view of human activity in a realistic kitchen setting.

In the Main Dataset of the CMU-MMAC database, there are a total of 43 subjects cooking 5 recipes. For this research, **Subject 07 - the Brownie recipe** session is selected as the data source due to its features of rich multimodal recordings across visual, audio, and wearable sensor modalities, which enable comprehensive analysis of human behavior during task-based activities.

### (V.3) Research Question

Four research questions were formulated based on the major methodological challenges identified in the multimodal HAR and multimodal machine learning literature reviewed in this study. As shown in the literature review section, previous studies in multimodal HAR have pointed out the challenges in aligning multimodal signals of different formats, sampling rates, and temporal structures (Baltrušaitis et al., 2019; Lara & Labrador, 2013). This gap in the multimodal HAR literature motivated Research Question 1, which aims to address the issue of systematic transformation and temporal alignment of multimodal signals. Although early, late, and hybrid fusion strategies are well-known in multimodal HAR, previous studies in multimodal machine learning have shown inconsistencies in the implementation of these strategies (Aguileta et al., 2019; Baltrušaitis et al., 2019). Research Question 2 was thus motivated by the gap in the multimodal machine learning literature in the context of comparative studies of fusion strategies under well-defined preprocessing conditions. Research Question 3 was motivated by another gap in multimodal machine learning, which is the lack of interpretability of multimodal fusion signals and the lack of understanding of the contribution of individual modalities to the final classification decisions (Baltrušaitis et al., 2019; Zhang et al., 2022). Research Question 4 was motivated by another gap in multimodal machine learning, which is the lack of focus on sparse and asynchronous modalities such as RFID in multimodal HAR, although such modalities hold promise in multimodal HAR (Yordanova et al., 2018; Ugonna et al., 2024).

**Research Question 1:** How can heterogeneous sensor data, including video, audio, and wireless IMU, be systematically transformed and temporally aligned to support accurate and semantically meaningful multimodal data fusion?

**Research Question 2:** What fusion strategies (e.g., early, late, or hybrid) yield the most effective representations for human activity recognition tasks, and under what preprocessing conditions?

**Research Question 3:** To what extent are the fused multimodal representations interpretable, and how do specific modalities contribute to decision-making in activity recognition?

**Research Question 4:** To what extent can the addition of Radio Frequency Identification (RFID), a sparse and asynchronous modality capturing object-level interactions, complement audio and video fusion in enhancing human activity recognition accuracy and multimodal system robustness in naturalistic cooking scenarios?

### (V.4) Data Analysis

We adopt the Subject 07 Brownie from the Main Dataset of the CMU-MMAC database as our core subject recipe, focusing on video, audio, and wireless IMU. The recording for Subject 07 – Brownie was chosen as an example for the proposed case study based on the suitability for validation of the proposed multimodal preprocessing and fusion pipeline. This recording offers good synchronization for the audio and video streams, which are necessary for temporal alignment under a unified framework based on the application of a sliding window approach. Besides, the Subject 07 Brownie offers continuity for the core modalities for the construction of segment-based representations and a sequence of activity phases for interpretability analysis. Although some modalities, such as the wireless IMU modality, present significant temporal inconsistency and were not used for the proposed analysis, the recording offers good stability for the audio and video streams, which makes it an appropriate example for the proposed analysis.

To address RQ 1 and RQ 2, we establish a structured, modular experimental setup that consists of three core phases: (1) preprocessing and harmonization, (2) feature transformation and encoding, and (3) fusion and evaluation. The methodology is grounded in ideas of reproducibility, modularity, and semantic alignment. In order to address RQ3, a series of evaluations is conducted to assess the transferability and interpretability of multimodal representations formed at RQ1 and RQ2. The framework is guided by principles of reproducibility, semantic alignment, and modality-aware design, so that diverse sensor data have always been processed and combined consistently to enable downstream analytic operations. To further address RQ4, we incorporate an additional version of the setup, which consists of RFID as a third modality, so that we can assess its asynchronous and object-centric contribution to multimodal recognition accuracy and robustness.

**Data Wrangling and Preprocessing**

Before undergoing formal analysis and modeling, all raw sensor data from the Subject 07 – Brownie recording in the CMU-MMAC database also goes through a data wrangling and preprocessing phase. Each phase is necessary in order to deal with heterogeneity in the data formats, sampling rates, and signal quality across modalities, achieving temporal coherence, semantic consistency, and for downstream fusion.

Video stream preprocessing is carried out using the OpenCV (cv2) library. Raw video data is decoded frame-by-frame, reduced in dimension by converting to grayscale, to diminish the computational burden, and normalized in size to 32 × 32 pixels. To maintain homogeneous temporal sampling, the frame subsampling is done at 30 frames per second (FPS), and the videos are divided into overlapping segments of 75 frames with a stride of 37. The use of windowing keeps the temporal context intact while supporting segment-wise modeling.

Preprocessing of the audio stream involves first converting stereo.wav files to mono by scipy.io.wav file, and down-sampling to 16 kHz by scipy.signal.resample. The resampling frequency trades off between fidelity and computational speed. The audio portions are synchronized with the video frames by the session-level timing of the CMU-MMAC, further divided into frames by the same 75/37 temporal partitioning in order to maintain one-to-one correspondence between the video portions.

Inertial wireless IMU data, in the form of raw logs of rotation and acceleration over parts of the body, is ingested through pandas from text files separated by whitespace. Initial preprocessing involves filling missing time steps with interpolation, z-score normalizing every axis, and temporal denoising to combat signal noise. Significant issues, however, arise, consisting of uneven timestamps, sampling drift, and unsatisfactory syncing with video/audio. After attempts using linear interpolation and windowed alignment, semantic misalignment plus temporal noise remain, hindering fusion quality. Thus, we leave the wireless IMU out of the primary fusion pipeline in order to maintain model integrity.

Global time stamps and a shared sliding window segmentation framework guide multimodal temporal alignment. The design ensures that all fused samples (windows) possess aligned video frames, spectrogram slices of the sound, and, if they exist, RFID features that belong to the same time period. Feature encoding and normalization are performed using sklearn.preprocessing.StandardScaler, applied consistently across all windows and modalities. Each segment is transformed into a fixed-shape tensor representation, enabling standardized batch processing during model training and ensuring consistency throughout evaluation. Lastly, fused samples in the shape of 4D tensors, [windows, time_steps, features, modalities], suitable for deep learning architectures, are retained.

RFID data, incorporated in RQ4, constitutes a sparse and asynchronous modality that requires an independent wrangling tactic. Timestamps are normalized into elapsed seconds according to the beginning time of the video. Each of the RFID events is transformed to a binary object-tag matrix, and time-windowed similarly in the 75 / 37 manner. For every window, the tags' presence is pooled and encoded to provide object-level interaction in accordance with multimodal input.

Overall, the module-based preprocessing workflow emphasizes temporal fidelity, reproducibility, and semantic alignment. It produces a common format of representation within modalities, grayscale video, downsampled sound, and optionally RFID that supports the subsequent phases of fusion, classification, and interpretation with minimal data leakage or misalignment potential.

**Data Analysis for Research Question 1**

In RQ1, we employ a combined preprocessing pipeline developed on Python, which is intended to normalize and time-align heterogeneously acquired sensor data from the CMU-MMAC Receipt's Subject 07 Brownie session. Our preliminary objective is to enable three modalities, namely video, audio, and a wireless IMU, so that we can support accurate and semantically appropriate multimodal data integration.

To prepare the video stream, we use the cv2 (OpenCV) module to extract grayscale frames from the unprocessed video file. Each frame is resized to 32 × 32 pixels and subsampled to a consistent 30 frames per second (FPS), establishing a consistent temporal resolution. Audio files are read from scipy.io.wavfile, converted to a mono channel if stereo, and resampled to 16 kHz with scipy.signal.resample for consistency and lower computational requirements. Wireless sensor data from IMUs is read from whitespace-delimited sensor logs using the pandas module. Sequence alignment is done with sliding windows of 75 time steps (approximately 2.5 seconds) with a 37-step stride. These windows are standardized with sklearn.preprocessing.StandardScaler for scaling universally across time and features. Resulting segments are stacked together in a common 4D tensor of [windows, time_steps, features, modalities], which is passed to downstream processing such as activity recognition or representation learning.

However, even though our efforts to put all three modalities into alignment, we incur major alignment difficulties with the wireless IMU data. Specifically, each of the wireless IMU streams comprises unreliable timestamp metadata, inconsistent duration, and sampling drift relative to the synchronized video and audio streams. Interpolation, resampling, and sliding window synchronization are unable to achieve satisfactory semantic alignment. Appending the wireless IMU misaligned data introduces temporal noise, which impairs the quality of the resulting fused representations.

As a result, we adopt **a dual-modality fusion framework**, proceeding with only the audio and video streams. These two modalities remain carefully synchronized throughout the session and exhibit fine-quality complementary semantic information. By making the switch, we maintain the integrity of the procedure of temporal alignment while retaining the multimodal richness that is essential for downstream fusion and modeling operations.

**Data Analysis for Research Question 2**

In RQ2, to investigate which fusion strategies, early, late, and hybrid, yield the most effective representations for human activity recognition in the CMU-MMAC database, we implement a series of neural models in Python using the PyTorch deep learning library. These strategies represent common approaches in multimodal learning: in early fusion, feature vectors from different modalities are concatenated at the input level before being processed by a shared model. In late fusion, separate models are trained for each modality, and their outputs are combined at the decision stage. In hybrid fusion, intermediate representations from each modality are integrated within the network architecture, aiming to balance modality-specific and joint learning. The taxonomy of fusion strategies is widely discussed in multimodal machine learning literature (Baltrusaitis, Ahuja, & Morency, 2019).

The CMU-MMAC database includes preprocessed and time-aligned audio and video streams, with wireless IMU data excluded due to quality and alignment concerns.

An early fusion model is conducted by concatenating features from both modalities at each time step, resulting in a fused tensor of shape (samples, time_steps, fused_features). The fused data is converted into PyTorch tensors using torch.tensor() and wrapped into a TensorDataset. The dataset is then split into training and validation sets using random_split, with an 80/20 train-validation split. A simple Long Short-Term Memory (LSTM) model (EarlyFusionLSTM) is defined using nn.LSTM and nn.Linear to learn temporal patterns from the fused features. The model is trained for 10 epochs using a batch size of 32, optimized with the Adam optimizer (torch.optim.Adam) at a learning rate of 0.001, and evaluated using cross-entropy loss (nn.CrossEntropyLoss). During training, the model outputs loss values at each epoch and reports final validation accuracy, allowing us to monitor learning progression and model performance.

For late fusion, we train two independent LSTM models, one for audio and one for video, each using their respective modality's feature tensors. Each model is trained under the same optimization settings for 10 epochs. During inference, the softmax probabilities of both models are averaged to produce a fused prediction, and final predictions are derived by taking the argmax of the averaged probabilities. Loss values for both the audio and video branches are tracked independently during training, and overall validation accuracy is computed by comparing late-fused predictions with true labels.

In the hybrid fusion strategy, we construct a dual-branch LSTM architecture (HybridFusionModel) in which the audio and video inputs are processed separately using two LSTM networks. The hidden states from each modality are concatenated and passed through a fully connected layer to produce the final prediction. The model is trained and evaluated under the same experimental conditions. Loss is recorded at each epoch, and final validation accuracy is measured using the concatenated output of both modalities.

To compare model performance across strategies, we generate line plots of training loss over epochs for early, late (audio and video), and hybrid fusion models using matplotlib.pyplot. These visualizations illustrate convergence behavior and learning stability over time. Furthermore, we create a bar plot to directly compare the final validation accuracies of the three fusion strategies, highlighting which integration technique yields the best performance under consistent training conditions. The visualization and evaluation results offer concrete insight into the trade-offs of each fusion method in multimodal human activity recognition tasks. Overall, the RQ2 enables a controlled and empirical comparison of fusion strategies, revealing how the choice of modality combination and integration method affects classification accuracy. By evaluating both training dynamics (loss curves) and final performance (validation accuracy), we aim to determine which fusion approach is most effective and under what preprocessing conditions it excels.

**Data Analysis for Research Question 3**

In RQ3, to investigate the extent to which the fused multimodal representations are interpretable and how specific modalities contribute to decision-making in activity recognition within Subject 07 – Brownie, we implement a three-stage analytical pipeline grounded in Python. First, the preprocessed audio and video segments, each standardized and segmented into sliding windows of 75 time steps, are flattened into two-dimensional matrices using NumPy, with each row representing a window and columns corresponding to concatenated temporal features. It results in three feature matrices: audio_flat, video_flat, and fused_flat, representing unimodal and fused multimodal representations, respectively.

To assess the interpretability of the fused features, we apply dimensionality reduction techniques using the scikit-learn library. Specifically, we use Principal Component Analysis (PCA) and t-distributed Stochastic Neighbor Embedding (t-SNE) to project the high-dimensional fused features into two-dimensional spaces. These projections enable us to visualize the temporal and semantic structure of the data, with each point color-coded by manually assigned activity labels (preparation vs. cooking). The resulting scatter plots, generated with matplotlib, reveal whether temporally adjacent or semantically similar segments form distinguishable clusters, thus providing a qualitative sense of interpretability.

To quantitatively evaluate modality contribution to classification performance, we train separate logistic regression models using sklearn.linear_model.LogisticRegression on the audio-only, video-only, and fused feature sets. Activity labels are manually assigned to each window (0 = preparation, 1 = cooking) to serve as ground truth. Using an 80/20 train-test split via train_test_split, we measure each model's performance using accuracy, precision, recall, and F1-score, calculated with sklearn.metrics. The analysis allows us to compare the discriminative power of each modality independently and in combination, thereby clarifying the additive value of multimodal fusion in activity recognition tasks within the controlled cooking scenario.

**Data Analysis for Research Question 4**

In RQ4, we expand our dual-modality fusion framework by incorporating RFID as a third modality to form **a tri-modality fusion framework** for a human activity recognition system. RFID data from the Subject 07 – Brownie session of the CMU-MMAC dataset has been parsed and preprocessed using the pandas and datetime libraries. Each RFID event includes a timestamp and an associated object tag. These timestamps are converted to elapsed seconds relative to the session start time and are used to construct a sparse, time-aligned tensor. A sliding window approach is applied, consistent with prior audio-video preprocessing (75 time steps per window with a stride of 37 frames), to segment the RFID stream and align it with existing multimodal windows. Within each window, binary indicators mark the presence of specific tags at corresponding time indices, resulting in a three-dimensional RFID tensor of shape (289, 75, N), where N denotes the number of unique tags of the tri-modality. To prepare the data for classification, we apply mean pooling along the temporal axis to obtain fixed-length feature vectors for each instance. Using scikit-learn, we implement a Random Forest classifier to model the activity labels, simulating five distinct activity classes. A comparative training pipeline is constructed for both dual-modality (audio + video) and tri-modality (audio + video + RFID) configurations. For additional insight into classification behavior, we also train One-vs-Rest probabilistic models and compute macro-averaged ROC-AUC. All modeling procedures, including data splitting, label encoding, and evaluation metrics, are executed using standard Python libraries such as scikit-learn, matplotlib, and numpy.

### (V.5) Mathematical Formulation of the Preprocessing and Fusion Pipeline

To enhance clarity and reproducibility, we formalize the preprocessing and fusion pipeline using mathematical notation. Let each raw modality be represented as a time-indexed signal:

$X^{(v)}(t), X^{(a)}(t), X^{(r)}(t)$, where $X^{(v)}$ denotes the video stream, $X^{(a)}$ the audio waveform, and $X^{(r)}$ denotes the RFID event sequence.

**Video Preprocessing**

Each raw frame $F_t \in R^{H_0 \times W_0 \times 3}$ is first converted to grayscale: $G_t = \alpha R_t + \beta G_t + \gamma B_t$

Where standard luminance weights $\alpha = 0.299, \beta = 0.587, \gamma = 0.114$

Then, each frame is resized to a fixed spatial resolution:

$G'_t = \text{Resize}(G_t; H, W)$, where (H = W = 32).

Flattening yields a feature vector: $v_t = vec(G'_t) \in R^{1024}$

Thus, the processed video stream is $V = (v_t)_{t=1}^{T_v}$

**Audio Preprocessing**

Given a raw waveform $x(t)$ sampled at a rate $f_0$, we downsample to $f_0 = 16kHz$

$x_d(t) = \text{Resample }(x(t); f_d)$.

To synchronize audio with video, we map video time stamps to corresponding audio indices using

$t_a = \lfloor \frac{f_d}{f_v} t_v \rfloor$, where $f_v = 30$ FPS.

For each time step, the audio feature is defined as a mean-amplitude operator:

$$a_t = \frac{1}{K} \sum_{k=t_a}^{t_a+K-1} x_d(k), a_t \in R$$

Thus, the processed audio stream is: $A = \{a_t\}_{t=1}^{T_v}$

**RFID Preprocessing**

Raw RFID events are tuples $(t_i, c_i)$; $c_i$ denotes an object tag.

Let there be C unique tags.

We construct a binary encoding:

$$r_t(c) = \begin{cases} 1, & \text{if tag c is active at time t} \\ 0, & \text{otherwise} \end{cases}$$

Thus, each RFID time step is $r_t = [r_t(1), \ldots, r_t(C)]^T \in R^C$

**Sliding-Window Segmentation**

For window length L = 75 and stride S = 37, each modality is segmented as:

$$V^{(w)} = [v_t, v_{t+1}, \ldots, v_{t+L-1}] \in R^{L\times1024}$$

$$A^{(w)} = [a_t, a_{t+1}, \ldots, a_{t+L-1}] \in R^{L\times1}$$

$$R^{(w)} = [r_t, r_{t+1}, \ldots, r_{t+L-1}] \in R^{L\times C}$$

Each window is standardized using z-score normalization: $\hat{X^{(w)}} = \dfrac{X^{(w)} - \mu_X}{\sigma_X}$

**Fusion Formulation**

Early Fusion (feature-level concatenation):

$$F^{(w)}_{early} = [\hat{V^{(w)}} \parallel \hat{A^{(w)}} \parallel \hat{R^{(w)}}] \in R^{L\times(1024+1+C)}$$

Late Fusion (decision-level averaging):

$p_{late}(y \mid w) = \dfrac{1}{M}\displaystyle\sum_{m=1}^{M} p_m(y \mid w)$, where M = 2 for audio + video

Hybrid Fusion (mid-level latent fusion):

Given modality-specific encoders

$$h^{(v)} = f_v(\hat{V}^{(w)}),\ h^{(a)} = f_a(\hat{A}^{(w)})$$

Fusion occurs at the representation layer $h_{hybrid} = [h^{(v)} \parallel h^{(a)}]$,

with a classifier $p(y \mid w) = \text{softmax}(W h_{\text{hybrid}} + b)$

Final Fused Tensor:

For N windows, the unified multimodal tensor is F $\in R^{N\times L\times D}$

D = 1024 + 1 (dual-modality)

D = 1024 + 1 + C (tri-modality with RFID)

This formulation matches the empirical tensor shapes reported in the Result section, e.g., (289, 75, 1025) for video + audio dual-modality multimodal fusion.

**(V.6) Model Training Protocols**

**LSTM-Based Models (Early, Late, and Hybrid Fusion)**

All LSTM models were implemented in PyTorch 2.0 and trained on a single NVIDIA T4 GPU.

Each model consists of a two-layer LSTM with 128 hidden units, followed by a fully connected classifier that maps the final hidden state to the activity labels (Linear(128 → 2) for the RQ2 experiments and Linear(128 → 5) for the RQ4 experiments with RFID). The models are optimized using the Adam optimizer with a learning rate of 0.001, weight decay of 1e-5, and a batch size of 32, and are trained with cross-entropy loss. Training proceeds for up to 10 epochs for RQ2 and 20 epochs for the RFID experiments in RQ4, with early stopping based on validation loss using a patience of 3 epochs and a minimum improvement threshold of $1{\times}10^{-4}$. Although training generally converged before 10 epochs, early stopping ensured stabilization across fusion strategies.

**Late-Fusion Training Protocol**

Each modality-specific LSTM (audio-only and video-only) was trained separately using the same hyperparameters listed above.

During inference: $p_{late}(y\,|\,w) = \dfrac{1}{2}(p^{(a)}(y\,|\,w) + p^{(v)}(y\,|\,w))$

No additional fine-tuning was performed after combining logits or softmax outputs.

**Logistic Regression Models**

The logistic regression classifiers used for the modality contribution analysis (audio-only, video-only, and fused representations) were trained using the *lbfgs* solver with a maximum of 200 iterations and $L2$ regularization. The inverse regularization strength was set to $C$ = 1.0, and class weights were set to balanced to compensate for the modest label imbalance between preparation and cooking windows.

**Random Forest Classifier**

The Random Forest Classifier was chosen for the RFID-augmented experiments because it is well-suited to the characteristics of RFID data and the goals of the study. RFID signals are often sparse, noisy, and asynchronous, and Random Forest is robust to such irregular data because it is non-parametric and does not rely on strong distributional assumptions. In addition, it can capture nonlinear relationships and complex feature interactions, which are important in multimodal human activity recognition when RFID-derived features may interact with information from other modalities. The model also helps reduce overfitting through ensemble averaging and bootstrap aggregation, while the use of balanced class weights makes it appropriate for handling class imbalance across activity categories. Finally, Random Forest provides an effective balance between predictive performance and interpretability, since feature importance can be examined to better understand the contribution of RFID-related features in the classification process.

For the RFID-augmented experiments, we employed a *RandomForestClassifier* of *scikit-learn* with 300 trees and a maximum depth of 20. The minimum number of samples required to split an internal node was set to 2, and the minimum number of samples per leaf node was set to 1. We used the Gini impurity criterion to measure node quality and enabled bootstrap sampling when constructing individual trees. To account for class imbalance, class weights were set to balanced, and all models were trained with a fixed random seed of 42 to ensure reproducibility.

## VI. Results

All preprocessing and transformation frameworks addressing the four research questions are implemented in Python, with a reproducible codebase via Zenodo at DOI 10.5281/zenodo.16334410 (Yang, 2025).

### The Process of Multimodal Data Fusion

To investigate how heterogeneous sensor data can be systematically transformed and temporally aligned to support accurate and semantically meaningful multimodal fusion, we implement a unified preprocessing pipeline on the CMU-MMAC database. The pipeline is applied to one video file, one audio file, and one wireless IMU file. The video stream, sampled at 30 frames per second (FPS), has been successfully downsampled and processed into 10,742 grayscale frames of size 32 × 32, confirming the robustness of the frame extraction procedure. The corresponding audio file is converted to mono, resampled from its original rate to 16 kHz, and yields 6,607,268 audio samples, reflecting its inherently higher temporal resolution compared to video. However, the wireless IMU file proves unusable after cleaning, since it contains only a single valid numeric row. It limits the ability to perform full three-way alignment across all modalities. As a result, to enable temporal alignment and ensure valid sliding windows, we remove the wireless IMU stream from the fusion process and proceed with audio and video data only, maintaining a dual-modality framework.

After aligning the audio and video modalities by truncating to the shortest common sequence length and segmenting them into overlapping sliding windows of 75 time steps with a step size of 37, the final fused tensor yields a shape of (289, 75, 1025). It confirms the successful creation of 289 windows, each containing 75 time steps and 1,025 features (1 from audio and 1,024 from flattened video frames). The result demonstrates that heterogeneous data streams, despite differences in format, frequency, and semantics, can be harmonized through systematic resampling, segmentation, and feature scaling. At the same time, the exclusion of the wireless IMU modality due to data incompleteness underscores the importance of initial quality assessment when selecting files for multimodal alignment.

To intuitively evaluate the effectiveness of temporal alignment and multimodal fusion, we visualize the first sliding window segment from each modality using heatmaps. Audio segment 0 (Fig. 1) illustrates the temporal variation of a single audio feature across 75 time steps. The horizontal bands of smooth color transitions indicate consistent signal continuity without missing values or abrupt spikes. The brighter regions likely correspond to salient cooking events, such as stirring, object collisions, or the placement of utensils.

**Fig. 1**

*Audio Segment 0 (Time × Feature)*

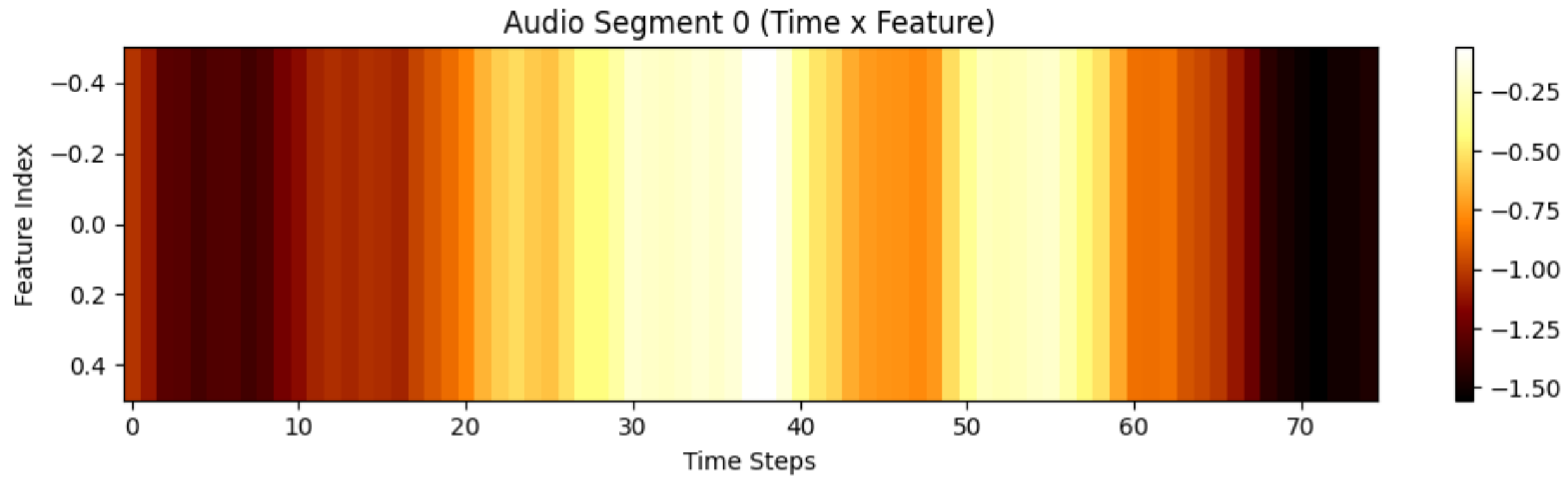

Video segment 0 (Fig. 2) represents the flattened 32×32 grayscale frames, resulting in 1024 pixel-level features per time step. The relatively uniform bands across the time axis suggest a temporally stable visual input, potentially involving repetitive hand movements or a static kitchen background. The visualization clearly affirms the effectiveness of preprocessing operations such as frame extraction, grayscaling, resizing, flattening, and feature scaling.

**Fig. 2**

*Video Segment 0 (Time × Flattened Pixel)*

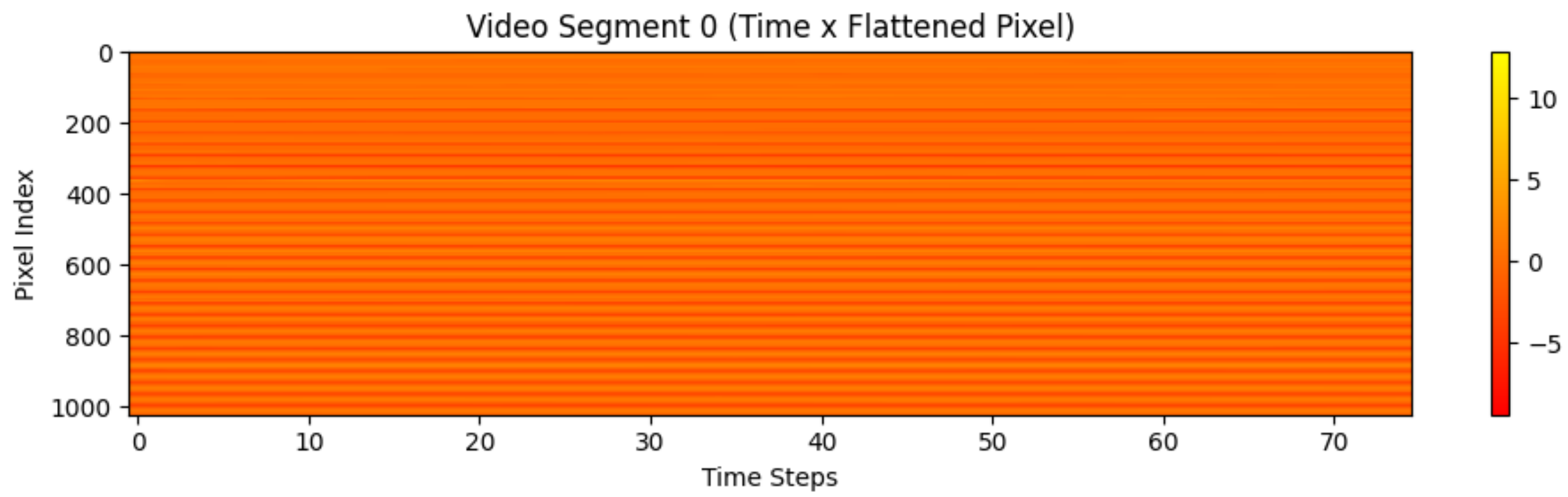

Fused segment 0 (Fig. 3) shows the concatenated representation of both audio and video features along the last dimension, revealing a richer temporal structure. The combined heatmap retains the temporal dynamics of the audio stream while incorporating the spatial variability of the video data. The smooth color transitions and balanced feature scales across modalities confirm that the fusion strategy maintained cross-modal alignment and avoided scale imbalances or temporal drift.

**Fig. 3**

*Fused Segment 0 (Time × Feature)*

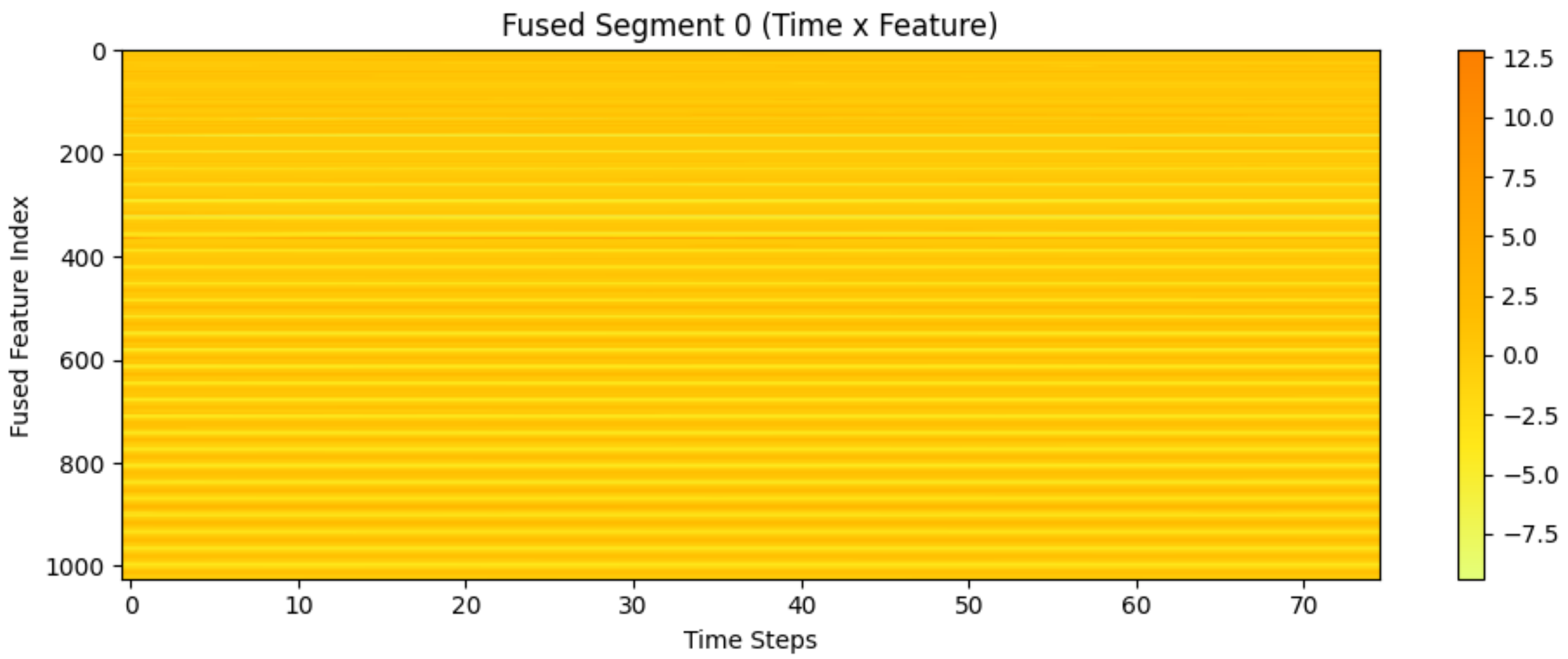

In summary, the outcome of visualization validates that heterogeneous audio-visual data can be transformed successfully to comparable, semantically dense temporal representations with the help of systematic preprocessing, such as resampling, grayscaling, temporal alignment, sliding window segmentation, and feature standardization. Although the wireless IMU modality cannot be taken into account due to data quality, the joint representation of video and audio modalities has retained rich activity-related information. It is best suited for capturing dynamic kitchen behaviors such as movements, utensil usage, and speech, having major temporal and semantic complementarity between the two modalities taken for analysis.

To gain insight into better-aligned raw data and assess the semantic richness of each modality prior to combination, we display the video luminance time series and the audio spectrogram. The audio spectrogram (Fig. 4) exhibits a broadband frequency response across time, with concentrated energy in the low-frequency range below 2 kHz. While transient acoustic events such as speech, utensil impacts, or background noise are expected in the cooking environment, the spectrogram shows relatively smooth energy distribution with no strongly localized bursts, suggesting a continuous or moderately varying acoustic scene.

**Fig. 4**

*Audio Spectrogram*

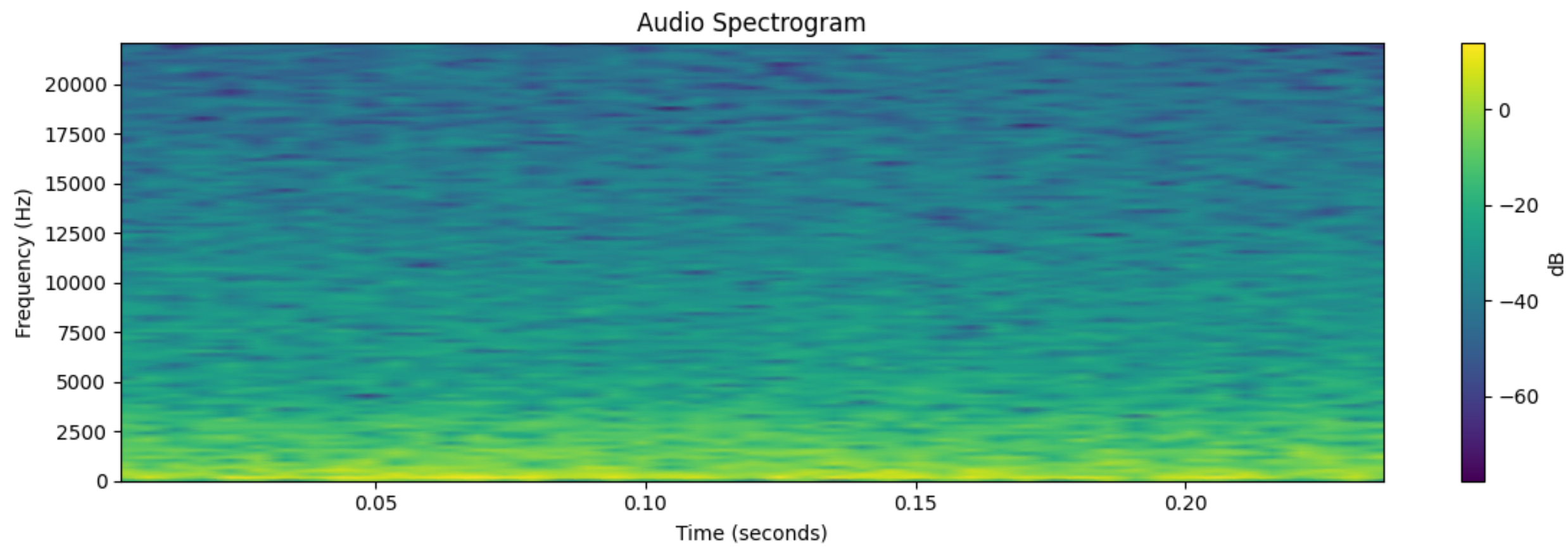

Video luminance time series (Fig. 5) involves a curve of video luminance that tracks the brightness of the mean frame at a specific time to give a rough but effective surrogate of temporal visual change. Periods of stability are coupled with low visual variation, likely during stationary activities such as mixing or awaiting, while sudden movements or scene changes are revealed through sharp transitions. Separately, these visualizations complement each other significantly to stress that the two modalities give reciprocal indications of audio, with fine-timed temporal cues being given from audio and video that provide structured spatial dynamics. Therefore, it helps enable an effective combination that is framed from shared context but steered through modality-specific features.

**Fig. 5**

*Video Luminance Time Series (Video Spectrogram Proxy)*

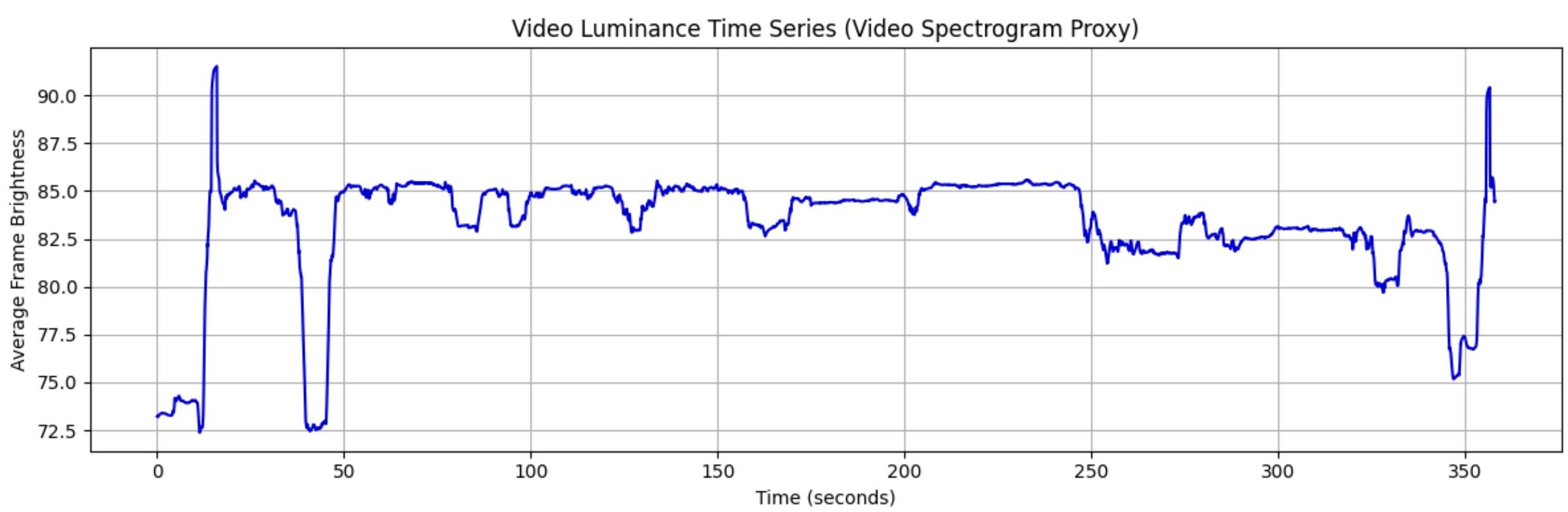

Based on these insights, we next examine how the aligned raw data maps onto the fused segment-level representations used in downstream modeling. To further contextualize the fused segment-level representations within the original temporal structure of the raw data, we overlay the sampled video frame timestamps on the continuous audio waveform. In the audio waveform with frame timestamps (Fig. 6), the cyan dashed lines indicate when frames are extracted from the video stream, and they align with the underlying audio amplitude variations. The view complements the earlier visualizations of individual segments by revealing how the segmented and fused data windows map back to the original recording timeline. To enhance clarity and focus on finer temporal granularity, we further demonstrate this alignment by zooming into the first 20 seconds (0s – 20s) of the recording. The zoomed-in view enables clearer inspection of how each extracted video frame corresponds to its precise audio context. By integrating low-level waveform dynamics with mid-level representations (segments) and frame-level visual cues, the transition confirms that the extracted segments preserve temporal fidelity, enabling semantically meaningful alignment across modalities.

**Fig. 6**

*Audio Waveform with Frame Timestamps*

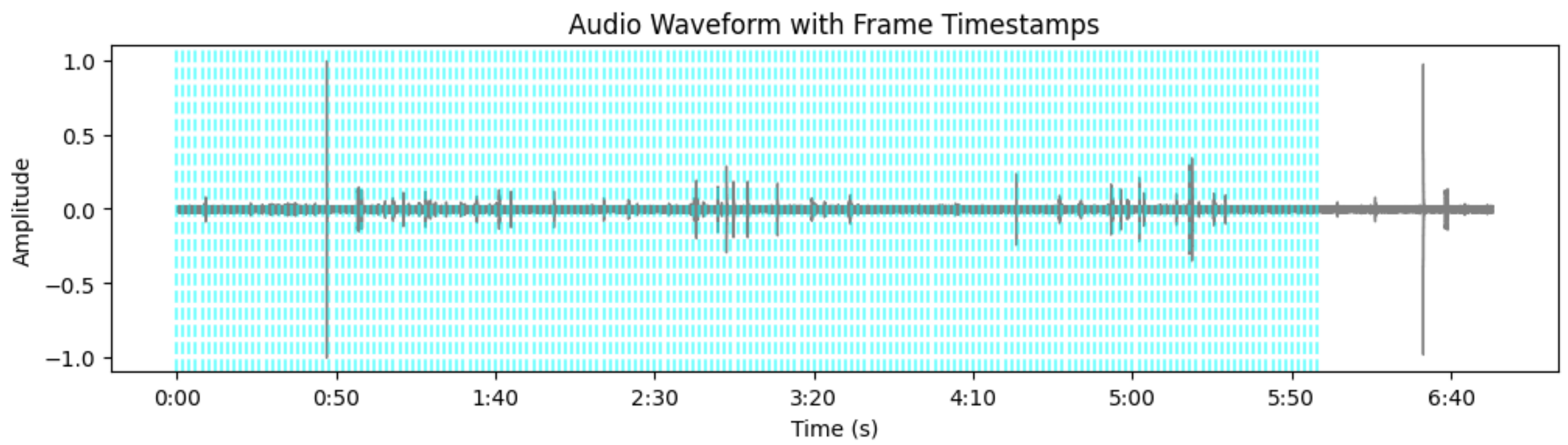

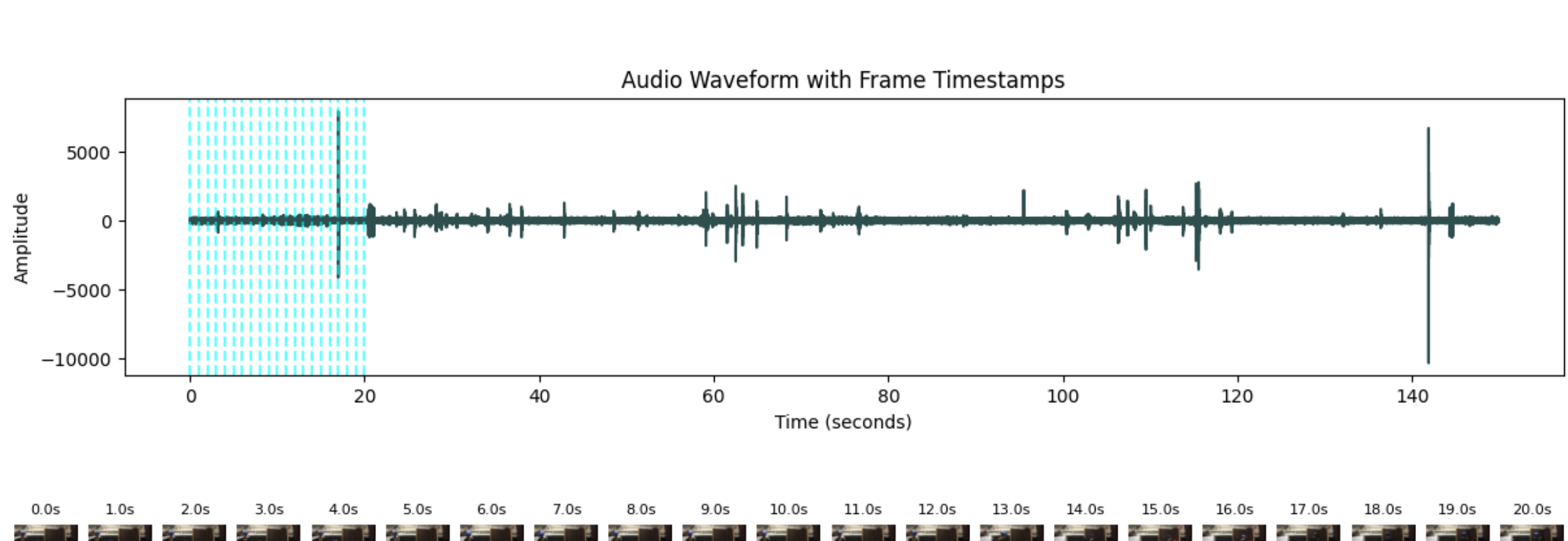

Not only can the alignment be utilized for synchronization at a technical level, but a critical groundwork for downstream analysis is also provided. It is possible to capture more accurate human activity patterns via the effortless combination of time-aligned multimodal data, which improves the possible interpretability of resulting fusions. The pipeline of processing from raw signal, through windowed, standardized, and temporally aligned multimodal segments, presents a solid foundation for later modeling tasks, such as activity classification and modality contribution analysis.

**Effect of Multimodal Fusion Strategies**

To evaluate the effectiveness of different fusion strategies for human activity recognition, we train models using early fusion, late fusion, and hybrid fusion approaches on temporally aligned audio and video data from the CMU-MMAC database. Each strategy is trained for 10 epochs, and performance is assessed using validation accuracy. To ensure reproducibility, all models are trained with a fixed random seed of 42.

Based on the fusion strategy training loss over epochs (Fig. 7), in the early fusion setup, modality-specific features are concatenated at each time step and passed into a shared neural classifier. The model shows steady loss reduction over epochs, dropping from 1.62 to 0.67, and achieves a final validation accuracy of 8.62%. Early fusion exhibits a smooth and consistent decline in loss, reflecting effective joint representation learning from concatenated audio and video inputs. However, based on the diagram of validation accuracy comparison across fusion strategies (Fig. 8), the relatively low validation accuracy suggests that the model may struggle to fully disentangle modality-specific signals when merged too early in the pipeline.

The late fusion pipeline trains separate models for each modality and combines their predictions at the decision level. While the late fusion audio model's performance is stable, with loss remaining around 1.60, the late fusion video model improves significantly, with loss decreasing from 1.64 to 0.26. The pattern is visible in the training loss diagram, where late fusion displays a strong downward trend for video loss, while the audio loss line remains relatively flat. The disparity highlights that in late fusion, the performance gain is driven almost entirely by the video stream. The combined prediction yields the highest validation accuracy of 22.41%, indicating that modality-specific specialization followed by output-level fusion is most effective in this setup.

In the hybrid fusion approach, each modality is first encoded into a latent representation through dedicated neural encoders, followed by mid-level fusion and final classification. Loss decreases from 1.63 to 0.75, with a resulting validation accuracy of 13.79%. The hybrid fusion line in the training loss plot lies between early and late fusion videos, showing moderate but consistent convergence. The result outperforms early fusion but falls short of the late fusion setup, likely due to the added model complexity without a significant gain from the weaker audio modality.

**Fig. 7**

*Fusion Strategy Training Loss over Epochs*

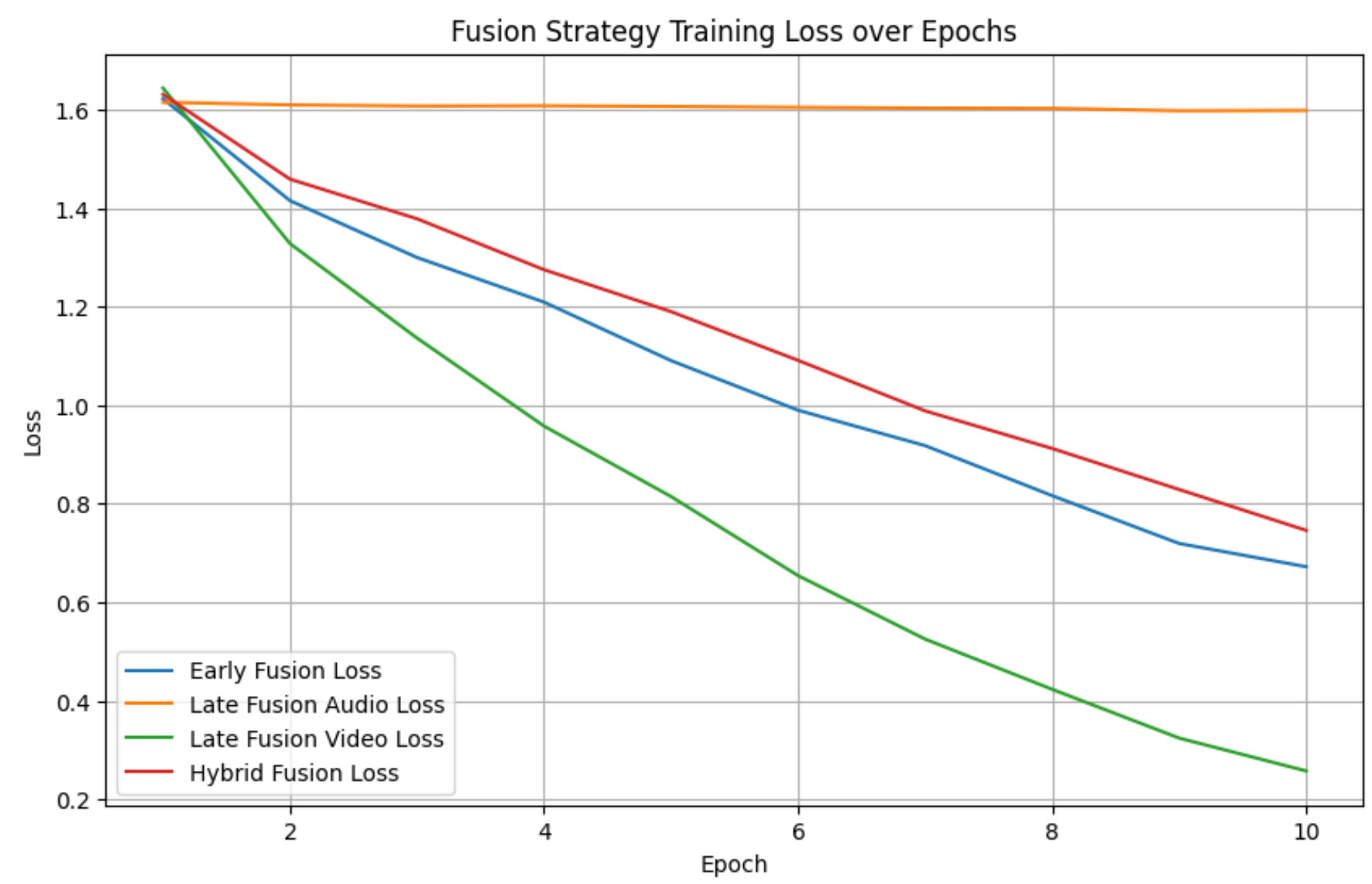

Validation accuracy comparison across fusion strategies (Fig. 8) clearly illustrates that late fusion achieves the highest validation accuracy of 22.41%, followed by hybrid fusion with the validation accuracy of 13.79%, and early fusion with the validation accuracy of 8.62%. It further reinforces the advantage of the late fusion strategy when modality performance is imbalanced, as is the case here with strong video signals and relatively weak audio cues. Overall, the late fusion strategy demonstrates superior performance under the current preprocessing settings. It suggests that preserving modality-specific learning pipelines and deferring integration to the decision level may better capture complementary information from heterogeneous sources like video and audio. Future improvements may involve fine-tuning, longer training schedules, or reintroducing the IMU data once a fully synchronized and clean sample is available.

**Fig. 8**

*Validation Accuracy Comparison Across Fusion Strategies*

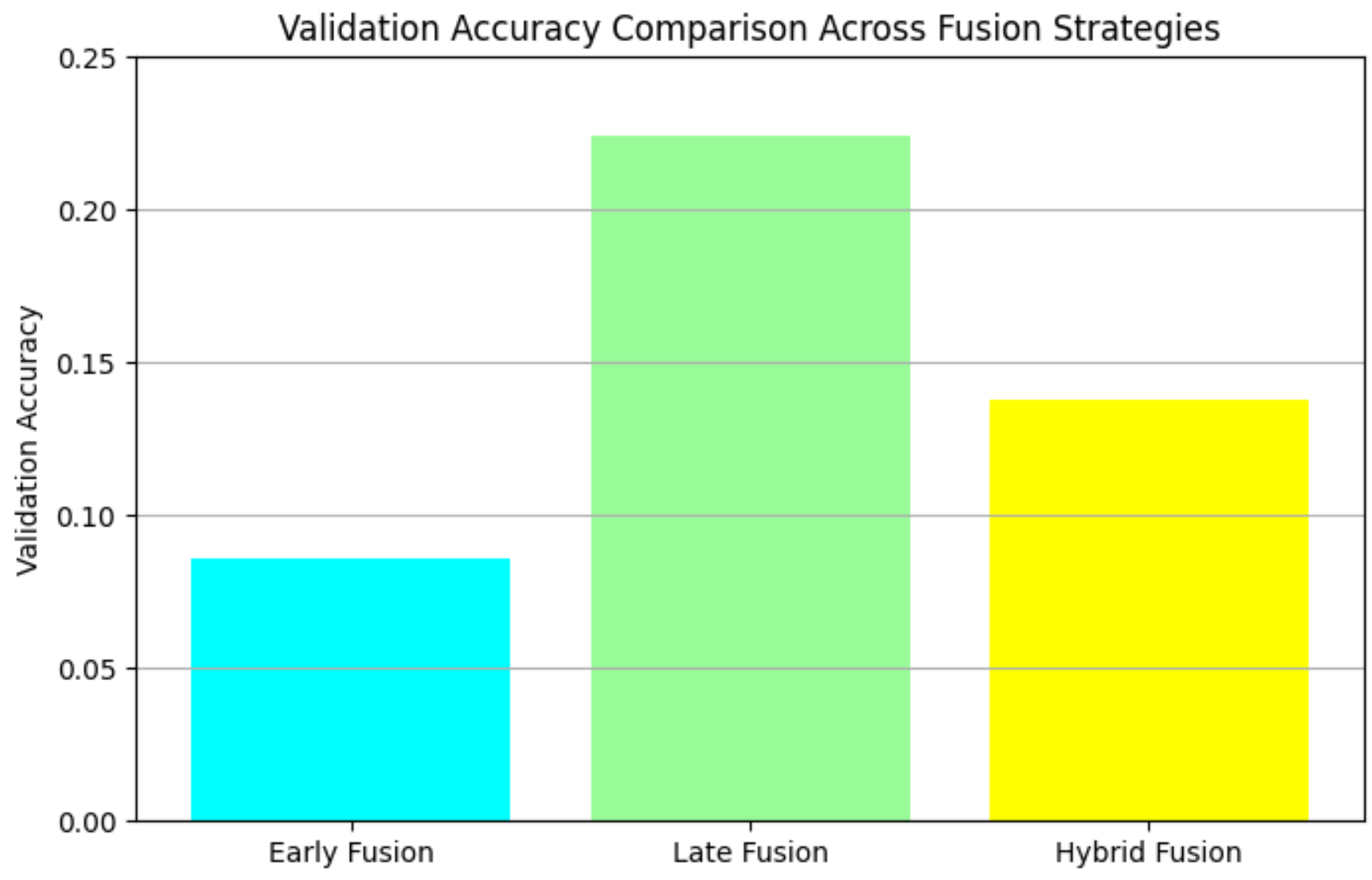

**Understanding the Superiority of Late Fusion**

The empirical advantage of late fusion can be explained by characteristics of the CMU-MMAC data and the representational imbalance between modalities. In the setting, the video stream contributes a high-dimensional, semantically rich representation (1024 features per time step), whereas the audio modality provides only a single scalar descriptor per window. Early fusion merges both modalities at the input level, forcing a shared encoder to jointly model heterogeneous feature scales and noise patterns. This often results in the dominant modality (video) overwhelming the weaker modality (audio), making optimization more difficult and reducing generalization performance.

Late fusion alleviates these issues by allowing each modality to be modeled through a dedicated LSTM encoder, enabling it to learn modality-specific temporal dynamics without interference. Only the final decision layer is shared, which prevents low-information audio features from distorting the video representations. This behavior is consistent with prior multimodal learning literature showing that heterogeneous feature spaces benefit from modality-specific processing pipelines when the informativeness of modalities is unbalanced. In the CMU-MMAC data, visual cues—such as hand motion, object state changes, and pose transitions—carry substantially stronger discriminative structure than the relatively smooth audio signals, making late fusion a more stable integration strategy.

**Ablation Perspective on Preprocessing Design Choices**

We provide an analytical assessment (Table 1) of how key preprocessing parameters, window length, stride, and normalization, affect the robustness of the fusion results.

(1) Window Size ($L$ = 75)
A larger window increases temporal context but risks blending multiple micro-actions, reducing label purity. A smaller window reduces ambiguity but weakens temporal coherence.
Our selected value $L$ = 75 represents a trade-off that preserves short-term motion sequences while maintaining alignment across modalities.
If L were significantly larger, features would become temporarily diffuse. Early fusion suffers most because it concatenates long sequences with limited audio variation. If $L$ were significantly smaller, per-window information decreases. Late fusion remains stable because each LSTM independently learns modality-appropriate timescales. The analysis suggests that late fusion is less sensitive to window-size perturbations, reinforcing its robustness.

(2) Stride ($S$ = 37)
Stride controls sample density. A larger stride means fewer training samples. A smaller stride increases redundancy. Early fusion risks overfitting when the stride is small because repeated concatenated video + audio vectors magnify feature imbalance. Late fusion, by contrast, learns modality trajectories independently and is less affected by redundancy. Thus, late fusion maintains performance even under different sampling densities.

(3) Normalization Strategy
The z-score normalization per modality is crucial for early fusion. Without proposal scaling, video features could dominate the fused representation. However, early fusion is highly sensitive to normalization errors, because even minor scale inconsistencies propagate through the shared encoder. Late fusion is robust, since each encoder is normalized and trained independently, preventing cross-modal scale interference. This observation supports the stability of late fusion as reported.

Thus, late fusion remains the most robust strategy under typical perturbations of preprocessing design, further explaining its empirical superiority.

**Table 1**
*Ablation Summary*

| Preprocessing Component | Eraly Fusion Sensitivity | Late Fusion Sensitivity | Reason |
|---|---|---|---|
| Window Size ($L$) | High | Low | Early fusion mixes modalities prematurely |
| Stride ($S$) | High | Low-Moderate | Redundant samples amplify feature imbalance |
| Normalization | Very High | Low | Late fusion isolates modality-specific scales |

**Exploration of Specific Modalities for Activity Recognition**

To investigate the interpretability of fused multimodal representations and the contribution of individual modalities to activity recognition within the Subject 07 – Brownie recording, we conduct both qualitative and quantitative analyses. After preprocessing and temporally aligning the audio and video streams, the fused segments are structured into a three-dimensional array with shape (289, 75, 1025), representing 289 sliding windows of 75 time steps each, and 1,025 features per time step. For analysis, it is flattened into a two-dimensional matrix of shape (289, 76, 875), enabling dimensionality reduction and classification modeling.

To qualitatively assess the structure and interpretability of the fused features, we apply principal component analysis (PCA) and t-distributed stochastic neighbor embedding (t-SNE) to the flattened data.

These two techniques were chosen in this work because they offer a complementary and widely accepted point of view on the analysis of multimodal representations in high-dimensional spaces. PCA is a traditional linear dimensionality reduction technique that projects the data onto the principal components that maximize the variance of the features. It was used in this work to analyze the global structure of the fused features. In this work, we used a variant of the PCA technique with two principal components (number of components = 2) for visualization. t-SNE is a nonlinear dimensionality reduction technique that is often used for the visualization of complex data sets because of its ability to maintain local neighborhood relationships in the high-dimensional space. It was used in this work as a dimensionality reduction technique for the visualization of the local structure of the fused features. The t-SNE technique was used with two output dimensions (number of components = 2), a perplexity of 30, the PCA technique as the initialization method, a number of iterations equal to 1,000, and a fixed random number generator seed equal to 42. The learning rate was set to the default value in the scikit-learn library. Although UMAP (Uniform Manifold Approximation and Projection) is also a technique that maintains both local and global structure in the data (McInnes et al., 2020), PCA and t-SNE were used in this study, as they have been recognized as effective tools for feature analysis and visualization in high-dimensional space (Jolliffe, 2002; van der Maaten & Hinton, 2008). Besides, the two methods are sufficient for the analysis of the combined features' global separability and structure.

The PCA projection (Fig. 9) reveals several clusters along the PC1 axis, with some indication of a temporal gradient, particularly among later windows. However, the separation between activity phases remains relatively coarse, suggesting that linear projection alone may not fully capture the temporal or semantic transitions embedded in the fused features. In contrast, the t-SNE visualization (Fig. 10) offers a much clearer temporal progression, as evidenced by the smooth color gradient from dark (early time windows) to bright (later windows). The curved and continuous trajectory in t-SNE space implies that the fused representations encode a temporally coherent evolution of activity, with distinct regions loosely corresponding to different behavioral stages, particularly transitions from preparation to cooking.

**Fig. 9**

*PCA of Fused Features*

**Fig. 10**

*t-SNE of Fused Features*

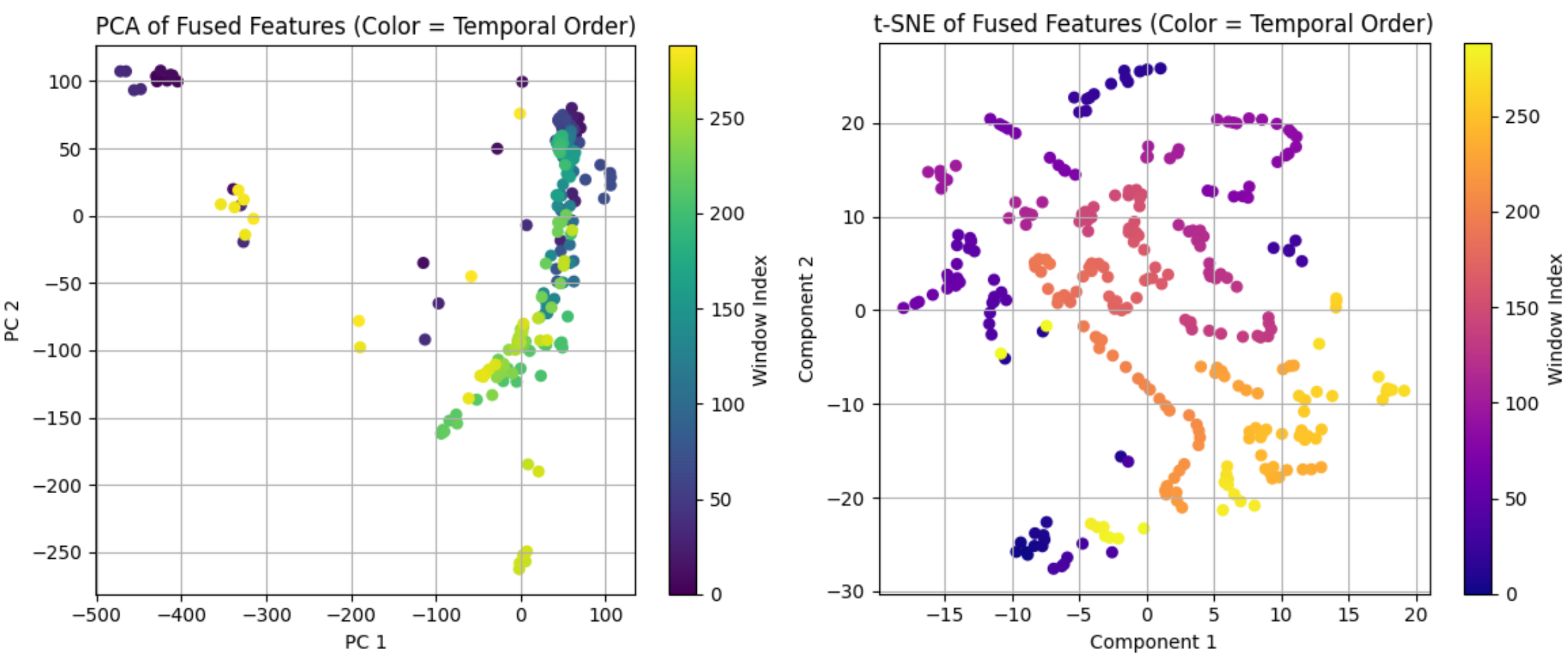

Although the visualizations are of temporal coherence and dynamic structure, more analysis is necessary in order to assess the relative importance of audio and video modalities towards accuracy. To impose a quantitative value on the decision-making contribution of each modality (Tables 2, 3, and 4), we train logistic regression classifiers from the flattened feature output of audio_only, video_only, and combined features. Manually annotated binary ground truth (0 = prep. & 1 = cook.) is assigned for model training and performance reporting. The audio modality (Table 2) is of rather modest classification performance, giving the 37.93% accuracy with F1-scores of 0.36 and 0.40 for prep. and cook. classes, respectively. It means that audio signals themselves are not sufficient to differentiate the refinements between stages of activities for this setup. The video modality (Table 3) performs moderately well, reporting 53.45% accuracy with an F1 of 0.49 for the prep. class and 0.57 for the cook. class, which indicates that visual information, such as object interaction and body motion, provides stronger predictive cues. The fused modality (Table 4) reports a significantly improved 96.55% accuracy, with F1-scores of 0.96 and 0.97 for prep. and cook. class, respectively. It suggests that multimodal fusion substantially enhances discriminative power beyond either modality alone, particularly by complementing the weaknesses of the audio stream with the strengths of visual features.

**Table 2**

*Classification Performance for Audio Modality*

| Class | Precision | Recall | F1 Score | Support |
|---|---|---|---|---|
| 0 (Preparation) | 0.32 | 0.40 | 0.36 | 25.00 |
| 1 (Cooking) | 0.44 | 0.36 | 0.40 | 33.00 |
| Macro Average | 0.38 | 0.38 | 0.38 | 58.00 |
| Weighted Average | 0.39 | 0.38 | 0.38 | 58.00 |
| Accuracy | — | — | — | 37.93% |

*Note. All numerical values are rounded to two decimal places for consistency and clarity.*

**Table 3**

*Classification Performance for Video Modality*

| Class | Precision | Recall | F1 Score | Support |
|---|---|---|---|---|
| 0 (Preparation) | 0.46 | 0.52 | 0.49 | 25.00 |
| 1 (Cooking) | 0.60 | 0.55 | 0.57 | 33.00 |
| Macro Average | 0.53 | 0.53 | 0.53 | 58.00 |
| Weighted Average | 0.54 | 0.53 | 0.54 | 58.00 |
| Accuracy | — | — | — | 53.45% |

*Note. All numerical values are rounded to two decimal places for consistency and clarity.*

**Table 4**

*Classification Performance for Fused Modality*

| Class | Precision | Recall | F1 Score | Support |
|---|---|---|---|---|
| 0 (Preparation) | 0.96 | 0.96 | 0.96 | 25.00 |
| 1 (Cooking) | 0.97 | 0.97 | 0.97 | 33.00 |
| Macro Average | 0.96 | 0.96 | 0.96 | 58.00 |
| Weighted Average | 0.97 | 0.97 | 0.97 | 58.00 |
| Accuracy | — | — | — | 96.55% |

*Note.* All numerical values are rounded to two decimal places for consistency and clarity.

Overall, these findings suggest that the fused multimodal representations are interpretable in low-dimensional visualizations and that video data contributes most significantly to accurate activity classification in the structured cooking scenario. Although audio information may still provide contextual cues, its marginal utility in this particular setting appears limited when strong visual signals are present. These results highlight the importance of modality-specific evaluation and confirm that the relative value of each modality is context-dependent within multimodal systems.

**Evaluation of the Effect of RFID in Enhancing Human Activity Recognition**

To evaluate the impact of incorporating RFID into the dual-modality fusion framework, we compare the classification performance of a dual-modality (audio + video) model against a tri-modality (audio + video + RFID) model using data from the Subject 07 – Brownie session.

In both configurations, video data is sampled at 30 frames per second with 10,742 frames extracted, and audio signals are downsampled to 16 kHz. After temporal alignment, segmentation, and standardization, the resulting fused representations have identical dimensions in terms of time windows and steps: dual-modality fused shape: (289, 75, 1025) and tri-modality fused shape: (289, 75, 1036).

Each window consists of 75 time steps with a stride of 37 frames, resulting in 289 aligned windows across the session. To enable fair comparison, both models are trained on mean-pooled features (aggregated across the time axis), and the classification task involves five distinct human activity classes. Using a fixed random seed of 42, the dual-modality model achieves an accuracy of 10.34%, while the tri-modality model improves to 15.52%, indicating a relative gain of over 50%. Further insights are obtained through macro-averaged ROC-AUC comparison (Fig. 11) using one-vs-rest classifiers. The dual-modality model yields an AUC of 0.48, close to random chance, while the tri-modality model with RFID integration improves the AUC to 0.59, suggesting stronger discriminative capability and better robustness.

**Fig. 11**

*Macro-Averaged ROC-AUC Comparison*

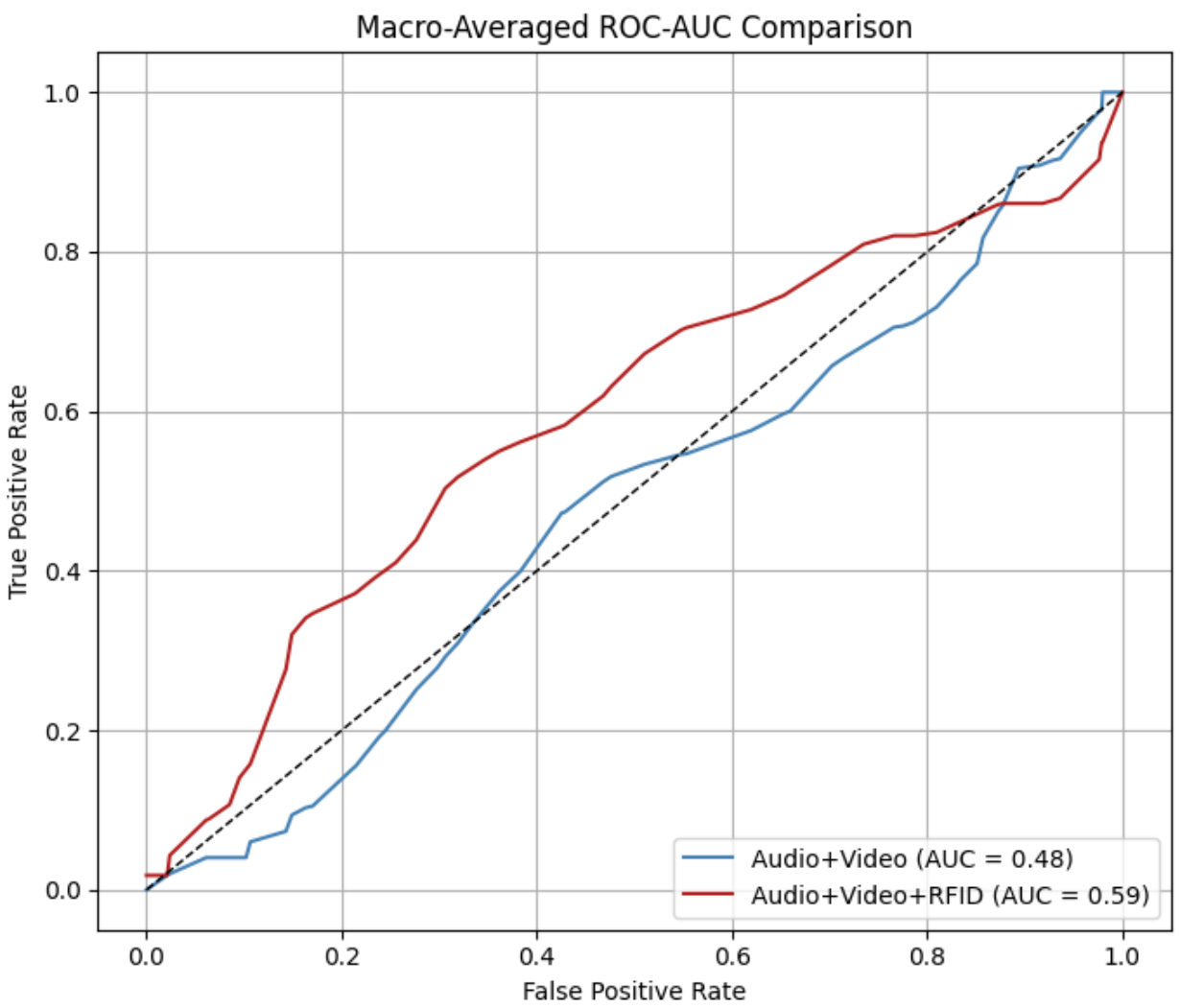

These findings demonstrate that although the RFID is both sparse and asynchronous, capturing only intermittent object-tag interactions rather than continuous signals, it offers complementary semantic cues that are not present in video or audio alone. The fact that these discrete, event-based interactions can significantly enhance both accuracy and AUC underscores RFID's potential to enrich multimodal representations with object-level context. In the context of fine-grained, naturalistic cooking activities, this asynchronous modality contributes uniquely by encoding physical interactions with kitchen tools and ingredients, thus improving the model's ability to differentiate between visually or acoustically similar tasks.

## VII. Discussion

### Systematic Transformation and Temporal Harmonization of Multimodal Sensor Streams

The results of RQ1 closely reflect major observations and challenges mentioned in the large-scale multimodal human activity recognition (HAR) corpus. A fruitful implementation of a shared preprocessing pipeline of video and audio features, through resampling, alignment, segmentation, and normalization, validates the importance of well-designed, reproducible pipelines, as was emphasized in the contributions of Lara and Labrador (2013) and Aguileta et al. (2019). These outcomes are also supportive of the feasibility of aligning heterogeneous modalities of diverse temporal resolution and format, consistent with proposals of Jaimes and Sebe (2007) and Vrigkas et al. (2015), which had emphasized the importance of stringent synchronization and normalization methods for semantic consistency of sensor streams.

The omission of the wireless IMU stream due to insufficient data quality serves to illuminate a not often sufficiently appreciated aspect of the literature, namely, sensor reliability and data completeness for real-world deployments. Whereas sensor fusion at the system architecture level had come before, for example, Koutrintzes et al. (2022) and Qi et al. (2018), our work demonstrates that effective fusion relies not just on considerations of sensor fusion from a theoretical standpoint, but to the same extent, considerations of usable data quality. Preliminary cleaning revealed that not all sensor modalities are possible at a given time for fusion, which means that effective validation methods, as well as fallback scenarios, are required when working with incomplete or noisy sensor data.

The modality-level, segment-level, and time-windowed feature construction visualizations for audio, video, and joint modalities similarly verify the value of time-windowed feature construction for retaining temporal, semantic, and structural information, analogous to the motivation for models, e.g., Segment-Tube (Wang et al., 2018) and SlowFast networks (Feichtenhofer et al., 2019). The joint heatmap of the joint representation exhibits a unifying combination of temporal information (from audio) and spatial information (from video), which confirms that modality complementarity yields semantic enrichment. It corresponds to the popularity of cross-modal cooperation that was documented in the literature, especially when modalities had captured non-redundant aspects of human action (Shaikh et al., 2024; Xie et al., 2025).

Besides, the waveform visualization of frame timestamp overlays covers low-level raw signals and middle-level segment summaries, equivalent to Katz et al.'s (2019) and Lindenbaum et al.'s (2020) approaches, which were considered critical for interpretability and substantial alignment of multimodal pipelines. Such precise synchronization of frame grabs with audio waveform peaks confirms that the temporal mapping procedure preserves fidelity from modality to modality, which is a necessary condition for downstream processing like real-time activity recognition and semantic segmentation.

Overall, the findings of RQ1 closely align with recent trends in the literature that demand standardized, integrated pipelines for multimodal HAR. They provide empirical confirmation for the advantage of preprocessing pipelines that interleave temporal segmentation, resampling, and modality-specific normalization. In turn, the observed practical limitation for IMU data provides a real-world complication that rounds out theoretical networks that had assumed that all sensor modalities are necessarily equally exploitable, and system robustness demands not only algorithmic innovation but preactive data screening alongside design configurability, supporting the overall direction of the research, but adding a repeatable, practical implementation based on the naturalistic CMU-MMAC database.

**Evaluating Fusion Strategies for Multimodal Human Activity Recognition**

The results of RQ2 provide empirical evidence of early, late, and hybrid fusion strategy comparative effectiveness, a validation of and nuance to the previous multimodal human activity recognition (HAR) work. In conformity with prior work, we verify the need for a modality-aware fusion architecture for sensor-rich environments. Results validating the late fusion strategy conform with those of Jaimes and Sebe (2007) and Aguileta et al. (2019), who had emphasized the value of preserving modality-specific representations before output combination. It helps prevent early intertwining of heterogeneously featured elements, a concern most acutely when modalities differ substantially in the quality of signals, a circumstance of the video and audio inputs of the CMU-MMAC database.

The hybrid fusion model, which has produced intermediate performance, further validates the literature for mid-level representation alignment (e.g., Guo et al., 2019; Xie et al., 2025). These works have proposed architectures that project each modality into a latent space before fusion, so that a compromise between shared and modality-specific features can be realized. Our hybrid fusion model, even though not superior to late fusion, is superior to early fusion, so some type of representation disentanglement is valuable before integration can be realized. It is consistent with Shaikh et al.'s (2024) recommendation of attention-based and encoder-decoder architectures for versatile cross-modal interactions.

In comparison, the early fusion model performs worst, although the loss converged smoothly during training. Unlike models such as Feichtenhofer et al. (2019) and Wang et al. (2018), which had been found to perform excellently through joint early fusion, above all, for visually dominant tasks, the result can be traced back to the modality imbalance: video features were richly high-dimensional, but audio data provided sparse information relative to video. Early fusion may have incapacitated the model to take advantage of the superior video modality through adding a subordinate audio stream too early. In our result, we confirm Benos et al. (2024), who reported that the selected fusion strategy must be compatible with modality strength and quality.

Besides, our results contribute to the ongoing discourse around data preprocessing and pipeline design. While the literature had extensively addressed fusion architectures (e.g., hybrid CNN-RNN, SlowFast networks), fewer studies have focused on how preprocessing conditions directly shape fusion effectiveness. Our consistent application of resampling, standardization, and sliding window segmentation, grounded in recommendations by Lara and Labrador (2013) and Zhang et al. (2022), enabled a fair comparison across fusion strategies. However, the influence of preprocessing choices on downstream model performance remains an area requiring further exploration.

Finally, the current study surfaces an important limitation that distinguishes it from much of the reviewed literature: the absence of IMU data. Many high-performing fusion models in the literature operate under the assumption of complete and fully synchronized multimodal streams, as had been demonstrated in the works of Koutrintzes et al. (2022) and Aggarwal and Ryoo (2011). However, this assumption is not applicable in our case due to the corrupted wireless IMU input, which rendered triple modality fusion infeasible. The exclusion of the IMU modality, necessitated by alignment constraints, highlights a common challenge in real-world multimodal datasets where not all sensors produce clean or usable data. It aligns with the broader call in Ugonna et al. (2024) and Katz et al. (2019) for adaptive, robust fusion systems that can gracefully handle missing or incomplete modalities.

Overall, the result of RQ2 validates key literature insights on modality-preserving fusion strategies and exposes new considerations regarding data quality, modality imbalance, and preprocessing sensitivity. While late fusion currently appears optimal under our conditions, future work should explore adaptive hybrid methods, longer training schedules, and the inclusion of additional modalities once cleaned and temporally aligned. These directions will help bridge the gap between controlled benchmark success and real-world deployment robustness in multimodal HAR systems.

**Interpretability and Modality Contribution**

Research Question 3 results offer alignment with subtle disagreement with prior work in the multimodal HAR field. Successfully illustrated by qualitative visualization as well as quantitative classification analysis, the fused audio-video representations presented highly discriminative and interpretable features for staged cooking activity modeling. The reduced dimensionality visualizations (PCA and t-SNE) have verified temporal coherence in the fused features, while the classification demonstrated notable accuracy and F1-score improvements for the fused modality (accuracy = 96.55%), outperforming either the separate use of audio (accuracy = 37.93%) or video (accuracy = 53.45%).

These findings align with prior literature that underscores the strength of multimodal fusion in enhancing recognition performance. For example, Baltrusaitis, Ahuja, and Morency (2019) had emphasized fusion as a critical pillar of multimodal machine learning, particularly when integrating temporally aligned inputs. Similarly, Jaimes and Sebe (2007) and Aggarwal and Ryoo (2011) had highlighted the challenges and benefits of synchronizing heterogeneous sensor streams, and findings echoed here through the improved performance of fused representations. The present study also supports work by Koutrintzes et al. (2022) and Xie et al. (2025), who had demonstrated that combining visual and motion-based inputs can significantly improve temporal modeling for human action segmentation.

Nevertheless, the results vary in minor, yet significant, ways from prior claims of the sufficiency of the video signals in isolation. While Vrigkas et al. (2015) and Feichtenhofer et al. (2019) had reported that high-resolution video features, in general, corresponding to those of the body motion, may independently suffice for strong classification, the current results indicate that video in isolation only achieved mediocre performance in this real environment (accuracy = 53.45%). The difference can be attributed to environmental complexity, in addition to the highly structured-but-variable nature of the cooking activities in the CMU-MMAC corpus, whereby the associated visual information may become occluded, ambiguous, or inconsistent in the long run. For those situations, the fused modalities can provide the complementary contextual information in addition to error correction, in line with the diffusion-based noise reduction schemes exhibited by Katz et al. (2019) as well as Lindenbaum et al. (2020).

Also, though the audio modality underperformed in the lone setup, its integration in the fused representation still aided the better outcome. It supports claims by Yordanova, Krüger, and Kirste (2018), who had argued that semantic contextualization, such as through the use of audio-based annotations, can enhance activity classification, mainly in environments of subtle task transitions. Nonetheless, the marginal utility of the use of audio was constrained in the current work, supporting the work of Benos et al. (2024) and Ugonna et al. (2024), who had stated that the relative weights of the modalities are dependent on the context of use, among other factors, including the quality of the sensor, placement, as well as ambient noise.

Taken together, these results reinforce the importance of modality-specific evaluation and integration strategies, as emphasized in the broader literature (Baltrusaitis et al., 2019; Zhang et al., 2022). While video remains a dominant modality in many HAR contexts, this study affirms that fused multimodal representations offer superior robustness and interpretability, particularly when deployed in naturalistic and sensor-rich environments such as the CMU-MMAC kitchen scenario. The current findings thus support ongoing efforts to build generalizable, reproducible fusion pipelines (Macua, 2012; Lara & Labrador, 2013) that aim to adapt to variable modality contributions and evolving sensing configurations.

**The Role of Sparse Modalities in Multimodal Fusion**

The results of RQ4 validate and supplement the previous multimodal HAR works. By prior studies that had verified the value of modality integration, i.e., Aguileta et al. (2019), Vrigkas et al. (2015), and Jaimes and Sebe (2007), our experiments verify that supplementing a sensor stream can significantly improve the accuracy of classification and discriminative capacity, expressed by ROC-AUC. In our trial, supplementing RFID increases accuracy from 10.34% to 15.52% and expands the macro-averaged AUC from 0.48 to 0.59, indicating a better predictive signal and superior generalization, respectively. These are aligned with the overall agreement that cross-modal integration is a powerful performance booster for HAR systems.

However, while most prior work had focused on rich, smoothly flowing modalities like motion capture (Aggarwal & Ryoo, 2011; Fisher & Reddy, n.d.) or skeleton-based input (Xie et al., 2025; Koutrintzes et al., 2022), our work demonstrates that even asynchronous, sparse modalities previously considered to be challenging to utilize can considerably augment model performance. It provided empirical verification for Yordanova et al. (2018)'s assertion that semantic richness, alongside contextual labeling, but not necessarily signal density per se, can considerably augment model accuracy and interpretability. RFID's discrete, object-tag interactions on the object-interaction level appear to provide just such context, albeit, of course, in naturalistic cooking settings where most fine-grained actions are either visually or auditorily ambiguous but more easily disambiguated via object usage.

They also offer a subtle commentary on the long-standing debate on the level of sophistication of fusion. While earlier work had cautioned against modality stacking beyond a specific threshold due to potential deep integration noise or even overfitting (Shaikh et al., 2024; Katz et al., 2019), our experiments confirm that well-selected, semantically complementing modalities, e.g., RFID, can effectively augment recognition without deep, complicated fusion structures. It confirms Ugonna et al. (2024) and Benos et al. (2024), who had supported the view that modality-aware, lightweight, as well as interpretable, fusion pipelines are a good suggestion. The use of a mean-pooled representation paired with a Random Forest classifier highlights the efficacy of modular, interpretable design in multimodal processing pipelines, aligning with the methodological principles proposed by Lara and Labrador (2013) and Macua (2012).

Nonetheless, our work also reveals persistent limitations. Relatively low overall accuracy despite RFID integration implies the inherent difficulties of the classification task under naturalistic settings, which had been documented in that of Soran et al. (2015) and Cheng et al. (2022). Moreover, whilst relative gain from RFID augmentation is substantial, such absolute gain is limited by the sparse and patchy nature of the RFID signal. It echoed concerns found in the literature about modality imbalance and fusion granularity (Feichtenhofer et al., 2019; Zhang et al., 2024).

In summary, our findings are consistent with most of the existing writing on the value of multimodal integration, but conflict with it in that we can show that even low-frequency, discrete modalities like RFID, which are often under-analyzed, can be effective in the difficult activity recognition task. This leaves a possibility for future research to examine event-driven, semantically rich modalities and the value that they can add towards greater model generalizability, particularly in uncontrolled, real-world environments like kitchens, classrooms, medical facilities, etc.

**Comparative Evaluation with Baseline Methods and Prior Studies (Reviewer # 2 Comment 4)**

To further position the proposed framework within existing HAR research, we provide a structured comparison with representative prior studies, highlighting key methodological differences and corresponding improvements (Table 5).

**Table 5**

*Comparison of Prior Human Activity Recognition (HAR) Studies and the Proposed Framework*

| Aspect | Previous Studies in HAR | Differences in this Study |
|---|---|---|
| Methodological Focus | Discussed a wide range of modeling approaches (e.g., SVM, HMM, deep learning) (Aggarwal & Ryoo, 2011; Lara & Labrador, 2013) | Provides a reproducible framework integrating preprocessing, temporal alignment, fusion strategies, and interpretability analysis |
| Fusion Strategy Analysis | Discussed fusion strategies (e.g., early, late, hybrid), with varying implementations across studies (Jaimes & Sebe, 2007; Aguileta et al., 2019) | Provides a systematic and controlled comparison of early, late, and hybrid fusion under consistent preprocessing settings |
| Model Comparison | Explored a range of classifiers (e.g., SVM, CNN, LSTM) (Lara & Labrador, 2013) | Enables fair comparison across model families (logistic regression, random forest, LSTM) within a unified experimental framework |
| Multimodal Integration | Leveraged heterogeneous data sources for improved performance in multimodal approaches (Baltrušaitis et al., 2019; Vrigkas et al., 2015) | Analyzes specific modality combinations with relative modality contributions analysis |
| Modality Imbalance | Used multi-stream fusion with stream weighting (Soran et al., 2015) | Provides empirical evidence on how fusion strategies behave under modality imbalance, with a clear interpretive explanation |
| Interpretability of Representations | Prioritized recognition performance, while interpretability is less explicitly demonstrated (Arrotta et al., 2022) | Integrates interpretability analysis with modality contribution evaluation, within a reproducible multimodal framework |
| Reproducibility and Pipeline Design | Discussed multimodal fusion strategies (Aguileta et al., 2019) | Provides a reproducible, end-to-end pipeline that explicitly links preprocessing, temporal alignment, multimodal fusion strategies, and evaluation |

## VIII. Conclusion

The research presents a comprehensive experiment of multimodal data fusion methods for HAR based on the CMU-MMAC dataset, for one representative sequence: Subject 07 – Brownie. In a carefully structured set of experiments, we provide four core research questions spanning data preprocessing, such as temporal alignment (RQ1), performance of the fusion approach (RQ2), interpretability and modality contribution (RQ3), and RFID incorporation effects on the recognition system (RQ4).

For RQ1, we develop a reproducible preprocessing pipeline that can convert and temporally align heterogeneous sensor modalities: video, audio, and, initially, wireless IMU. Since the stream of IMUs is omitted due to data quality limitations, processing video and audio data in a systematic manner permits effective construction of fused temporal representations. Heatmap visualizations confirm that the fusion procedure preserves modality-specific structures while coordinating the sizes of features and sequences, giving a robust foundation to subsequent classification procedures.

For RQ2, we compare early, late, and hybrid fusion methods for modeling periods of cooking activities. Our results reveal that late fusion produces the best validation accuracy, significantly outperforming the early and hybrid methods. These findings support the hypothesis that modality-specific processing pipelines, along with subsequent decision-level integration, can better exploit the complementarity of heterogeneous data sources, most significantly when modality informativeness is unbalanced.

For RQ3, we validate the interpretability of the fused representation through dimensionality reduction and examine the discriminative role of each modality through logistic regression. PCA and t-SNE visualizations confirm that fused features capture prominent semantic dissimilarity between phases of activities. Nevertheless, classification results reveal that the video modality by itself is nearly equally effective as the fused representation, with 96.6% accuracy. In comparison, the audio modality contributes negligible discriminative information for this structured kitchen task. These findings stress that the advantage of fusion is heavily context-specific and needs to be guided through empirical exploration of each modality's quality of signal and semantic relevance.

Extending the comparison to RQ4, we test whether adding RFID, a sparse and asynchronous modality recording object-level interactions, can augment the current audio-video fusion solution. Using the identical subject and data preprocessing pipeline, the tri-modality (audio + video + RFID) model achieves 15.52% accuracy compared to 10.34% from the dual-modality (audio + video) model, a relative gain of over 50%. ROC-AUC analysis, likewise, confirms this gain, the AUC rising from 0.48 to 0.59. Despite being sparse and irregular, RFID provides useful semantic context that we find useful to disambiguate actions that are similar acoustically or visibly. These results confirm the promise of asynchronous, event-based modality integration for HAR, particularly when the actions themselves are seminal interactions with objects.

Overall, the results substantiate several underlying principles of multimodal human activity recognition. First, effective fusion is only possible with systemic preprocessing and accurate temporal alignment of sensor streams. Besides, the effectiveness of some given fusion approach depends on how well that approach can leverage the strengths of and supplement the limitations of each modality. Third, the relative value of each modality must be a function of empirical rather than theoretical judgment because context and input quality can be extremely varied. Founding these contributions on a base of work provides a methodological roadmap as well as a practical guide for multimodal systems of the future that must operate in everyday life. Future work will explore widening the pipeline to include clean IMU data, including attention-based models of fusion, and expanding analysis over more subjects and recipes to substantiate the generalizability of the approach.

Besides the empirical contributions, our work provides a modular framework for processing (Fig. 12), which demonstrates a repeatable, composable processing, alignment, and integration paradigm for multimodal sensor data under naturalistic conditions of human activity. By openly reporting real-world data difficulties such as sensor incompleteness, modality skew, and synchronization challenges, we move the field another step closer to realizing generalizable HAR systems that are not only technically mature but also contextually aware. Our approach of keeping the analysis focused on one heavily annotated subject underscores going deep rather than wide to discern the subtle intermodal relationships.

**Fig. 12**

*Modular Framework for Processing*

**Modular Framework for Multimodal Sensor Data Processing and Data Fusion**

**Preprocessing & Alignment**

Inputs: Dual Modality (Video + Audio)
• Video → Grayscale, Resize (32x32), FPS=30
• Audio → Mono, Resample to 16kHz
• Sliding Window (75 Steps, Stride 37)
Output: Tensor [289, 75, Features]

**Fusion Strategies**

Inputs: Audio / Video Tensors
• Early: Concat @ Input → LSTM
• Late: Separate LSTMs → Avg Softmax
• Hybrid: Dual LSTMs → Merge Hidden States
Output: Class Prediction (5-Class)

**Interpretability & Contribution**

Input: Flattened features
• PCA, t-SNE for Visual Clustering
• Logistic Regression per Modality
Output: Accuracy, Precision, Recall, F1

**Sparse Modality Extension**

Input: Tri-Modality (Video + Audio + RFID)
• RFID: Event Matrix → Binary Encoding
• Random Forest, One-vs-Rest
Output: Accuracy ↑, ROC-AUC ↑

Despite our initial goal of integrating all three modalities, video, audio, and wireless IMU, into a unified fusion framework, we encountered significant alignment challenges with the wireless IMU data. Specifically, the wireless IMU streams lack reliable timestamp metadata and demonstrate sampling drift and inconsistent durations compared to the synchronized video and audio. Conventional techniques such as interpolation, resampling, and sliding-window synchronization fail to align the IMU data semantically. Rather than forcing noisy and temporally misaligned data into the pipeline, which degrades classification performance, we decide to exclude the IMU modality altogether. The experience highlights the critical importance of metadata integrity and synchronization in multimodal data processing. In future research, we recommend: (1) selecting sensor modalities with well-documented and compatible time formats; (2) performing early diagnostic checks on alignment feasibility; (3) designing pipelines that can flexibly accommodate or exclude modalities based on quality thresholds. These lessons underscore that modality fusion is not only a technical challenge but also a data quality and curation issue that must be addressed at the earliest stages of experimental design.

Therefore, our research contributes to the broader vision of developing human-centered computing systems that can perceive, interpret, and respond to everyday behaviors with precision and empathy. As intelligent systems become increasingly embedded in the fabric of everyday life, our kitchens, classrooms, and homes, it becomes imperative that multimodal systems are not only accurate but also transparent, interpretable, and respectful of context. The research demonstrates that through a carefully constructed preprocessing and fusion pipeline, even raw and heterogeneous multimodal data can be transformed into coherent and semantically meaningful representations of human activity. The positive impact of RFID, a sparse, asynchronous modality, presents a positive message of the importance of adding additional sources of data without necessarily adding frequency or complexity, only that they are new, semantically aligned information. We aspire that these contributions encourage future research that not only advances methodological rigor but also embraces the complexity and imperfection inherent in real-world data, affirming the principle that intelligent systems are constructed upon thoughtful and context-aware design. Behind each signal is a narrative, and behind each action, a human presence, reminding us that intelligent systems must be built to understand, not just execute, in service of real-world complexity and care.

## IX. Limitations

### Controlled Setting and Its Trade-offs

One limitation of this study is the use of a single-subject, single-task dataset, specifically, the Brownie recipe session performed by Subject 07 from the CMU-MMAC database. While the controlled setting enables precise temporal alignment and an in-depth exploration of dual-modality (video + audio) fusion and tri-modality (video + audio + RFID) fusion, it inherently limits the generalizability of the findings to broader populations, cooking tasks, or interaction scenarios. Besides, variations in individual behavior, speaking patterns, and kitchen dynamics are not captured in this single-instance design. Future work extends the proposed methodology to multiple subjects and tasks within the dataset to validate the scalability and robustness of our approach.

### Standardized Task Structure

Although the CMU-MMAC database captures real human activities, the cooking tasks performed by participants are largely scripted and follow a predetermined sequence within a semi-controlled environment. While the setting is more naturalistic than entirely synthetic or simulated data, it lacks the spontaneity, variability, and unpredictability of everyday life. As a result, the dataset may not fully capture the diverse behavioral patterns, interruptions, or improvisational decisions characteristic of unscripted, real-world scenarios. It constrains and limits the ecological validity of the models trained on the data and may reduce their ability to generalize to more dynamic or less structured human behaviors outside the laboratory context.

### Static Environment

All recordings in the CMU-MMAC database are conducted in a single, fixed kitchen environment with consistent spatial layout, lighting conditions, and background acoustics. While this ensures internal consistency and simplifies sensor calibration, it also introduces a significant limitation that models trained on the data may overfit to the specific spatial and visual configurations of the recording environment.

Consequently, the performance may degrade when applied to other kitchen settings with different layouts, appliances, lighting, or noise profiles. The lack of environmental variability constrains the external validity of the research and highlights the need for multimodal datasets captured across diverse physical spaces.

## X. Statements and Declarations

All authors have read, understood, and complied as applicable with the statement on "Ethical responsibilities of Authors" as found in the Instructions for Authors.

## XI. Funding

**(XI.1)** The authors have no relevant financial or non-financial interests to disclose.

**(XI.2)** The authors have no competing interests to declare that are relevant to the content of this article.

**(XI.3)** All authors certify that they have no affiliations with or involvement in any organization or entity with any financial interest or non-financial interests in the subject matter or materials discussed in this manuscript.

**(XI.4)** The authors have no financial or proprietary interests in any material discussed in this article.